\documentclass[12pt, onecolumn, draftclsnofoot]{IEEEtran}%

\usepackage[T1]{fontenc}
\usepackage{amsmath,amssymb,amsfonts,mathrsfs,bm}
\usepackage{mathtools}
\usepackage{amsthm}
\usepackage{cite}
\usepackage[shortlabels]{enumitem}
\usepackage{enumitem}
\usepackage{graphicx}
\usepackage{epstopdf}
\usepackage{url}
\usepackage{colortbl}
\usepackage{multirow}
\usepackage[table,dvipsnames]{xcolor}
\usepackage[normalem]{ulem}
\usepackage{array}
\newcolumntype{L}[1]{>{\raggedright\let\newline\\\arraybackslash\hspace{0pt}}m{#1}}
\newcolumntype{C}[1]{>{\centering\let\newline\\\arraybackslash\hspace{0pt}}m{#1}}
\newcolumntype{R}[1]{>{\raggedleft\let\newline\\\arraybackslash\hspace{0pt}}m{#1}}

\makeatletter
\let\MYcaption\@makecaption
\makeatother
\usepackage[font=footnotesize]{subcaption}
\makeatletter
\let\@makecaption\MYcaption
\makeatother

\usepackage{xparse}



\usepackage{glossaries}
\newacronym{wrt}{w.r.t.}{with respect to}
\newacronym{RHS}{R.H.S.}{right-hand side}
\newacronym{LHS}{L.H.S.}{left-hand side}
\newacronym{iid}{i.i.d.}{independent and identically distributed}

\usepackage{float}

\ifx\notloadhyperref\undefined
	\ifx\loadbibentry\undefined
		\usepackage[hidelinks,hypertexnames=false]{hyperref} 
	\else
		\usepackage{bibentry}
		\makeatletter\let\saved@bibitem\@bibitem\makeatother
		\usepackage[hidelinks,hypertexnames=false]{hyperref}
		\makeatletter\let\@bibitem\saved@bibitem\makeatother
	\fi
\else
	\relax
\fi

\usepackage[capitalize]{cleveref}
\crefname{equation}{}{}
\Crefname{equation}{}{}
\crefname{claim}{claim}{claims}
\crefname{step}{step}{steps}
\crefname{line}{line}{lines}
\crefname{condition}{condition}{conditions}
\crefname{dmath}{}{}
\crefname{dseries}{}{}
\crefname{dgroup}{}{}
\crefname{Theorem}{Theorem}{Theorems}
\crefname{Corollary}{Corollary}{Corollaries}
\crefname{Proposition}{Proposition}{Propositions}
\crefname{Lemma}{Lemma}{Lemmas}
\crefname{Definition}{Definition}{Definitions}
\crefname{Example}{Example}{Examples}
\crefname{Assumption}{Assumption}{Assumptions}
\crefname{Remark}{Remark}{Remarks}
\crefname{Rem}{Remark}{Remarks}
\crefname{remarks}{Remarks}{Remarks}
\crefname{Exercise}{Exercise}{Exercises}
\crefname{Theorem_A}{Theorem}{Theorems}
\crefname{Corollary_A}{Corollary}{Corollaries}
\crefname{Proposition_A}{Proposition}{Propositions}
\crefname{Lemma_A}{Lemma}{Lemmas}
\crefname{Definition_A}{Definition}{Definitions}

\usepackage{algorithm,algorithmic}

\ifx\loadbreqn\undefined
	\relax
\else
	\usepackage{breqn} 
\fi

\ifx\renewtheorem\undefined
\ifx\useTheoremCounter\undefined
\newtheorem{Theorem}{Theorem}
\newtheorem{Corollary}{Corollary}
\newtheorem{Proposition}{Proposition}
\newtheorem{Lemma}{Lemma}
\else
\newtheorem{Theorem}{Theorem}
\newtheorem{Corollary}[theorem]{Corollary}
\newtheorem{Proposition}[theorem]{Proposition}
\fi

\newtheorem{Definition}{Definition}
\newtheorem{Example}{Example}

\newtheorem{Assumption}{Assumption}

\fi

\theoremstyle{remark}

\theoremstyle{plain}

\newcommand{\Real}{\mathbb{R}}

\newcommand{\calF}{\mathcal{F}}

\newcommand{\bbZ}{\mathbb{Z}}

\DeclareSymbolFont{bsfletters}{OT1}{cmss}{bx}{n}
\DeclareSymbolFont{ssfletters}{OT1}{cmss}{m}{n}
\DeclareMathSymbol{\bsfGamma}{0}{bsfletters}{'000}
\DeclareMathSymbol{\ssfGamma}{0}{ssfletters}{'000}
\DeclareMathSymbol{\bsfDelta}{0}{bsfletters}{'001}
\DeclareMathSymbol{\ssfDelta}{0}{ssfletters}{'001}
\DeclareMathSymbol{\bsfTheta}{0}{bsfletters}{'002}
\DeclareMathSymbol{\ssfTheta}{0}{ssfletters}{'002}
\DeclareMathSymbol{\bsfLambda}{0}{bsfletters}{'003}
\DeclareMathSymbol{\ssfLambda}{0}{ssfletters}{'003}
\DeclareMathSymbol{\bsfXi}{0}{bsfletters}{'004}
\DeclareMathSymbol{\ssfXi}{0}{ssfletters}{'004}
\DeclareMathSymbol{\bsfPi}{0}{bsfletters}{'005}
\DeclareMathSymbol{\ssfPi}{0}{ssfletters}{'005}
\DeclareMathSymbol{\bsfSigma}{0}{bsfletters}{'006}
\DeclareMathSymbol{\ssfSigma}{0}{ssfletters}{'006}
\DeclareMathSymbol{\bsfUpsilon}{0}{bsfletters}{'007}
\DeclareMathSymbol{\ssfUpsilon}{0}{ssfletters}{'007}
\DeclareMathSymbol{\bsfPhi}{0}{bsfletters}{'010}
\DeclareMathSymbol{\ssfPhi}{0}{ssfletters}{'010}
\DeclareMathSymbol{\bsfPsi}{0}{bsfletters}{'011}
\DeclareMathSymbol{\ssfPsi}{0}{ssfletters}{'011}
\DeclareMathSymbol{\bsfOmega}{0}{bsfletters}{'012}
\DeclareMathSymbol{\ssfOmega}{0}{ssfletters}{'012}

\DeclareMathOperator{\st}{s.t.}

\DeclareMathOperator{\sinc}{sinc}
\DeclareMathOperator{\rank}{rank}

\DeclarePairedDelimiter\abs{\lvert}{\rvert}
\DeclarePairedDelimiter\parens{(}{)}

\newcommand{\qednew}{\nobreak \ifvmode \relax \else
      \ifdim\lastskip<1.5em \hskip-\lastskip
      \hskip1.5em plus0em minus0.5em \fi \nobreak
      \vrule height0.75em width0.5em depth0.25em\fi}

\newcommand{\nn}{\nonumber\\}

\newcommand{\ud}{\mathrm{d}}

\newcommand{\norm}[1]{{\left\lVert{#1}\right\rVert}}

\DeclareDocumentCommand \ifcond {m m} {%
	{#1} %
	\IfValueT{#2}{\, \middle|\, {#2}}%
}

\DeclareDocumentCommand \P {e{_} g >{\SplitArgument{ 1 }{ @| }}d() g } {%
	\mathbb{P}%
	\IfValueTF{#1}{_{#1}}
		{\IfValueT{#2}{_{#2}}}%
	\IfValueT{#3}{\left(\ifcond#3}%
	\IfValueT{#4}{\, \middle|\, {#4}}%
	\IfValueT{#3}{\right)}%
}

\DeclareDocumentCommand \E {e{_} g >{\SplitArgument{ 1 }{ @| }}o g } {%
	\mathbb{E}%
	\IfValueTF{#1}{_{#1}}
		{\IfValueT{#2}{_{#2}}}%
	\IfValueT{#3}{\left[\ifcond#3}%
	\IfValueT{#4}{\, \middle|\, {#4}}%
	\IfValueT{#3}{\right]}%
}

\definecolor{gray90}{gray}{0.9}

\ifx\nohighlights\undefined

	\newcommand{\msout}[1]{\text{\color{green} \sout{\ensuremath{#1}}}}
	\newcommand{\del}[1]{{\color{green}\ifmmode \msout{#1}\else\sout{#1}\fi}}
\else

	\newcommand{\msout}[1]{#1}
	\newcommand{\del}[1]{#1}
\fi

\newcommand{\hide}[1]{}

\renewcommand{\figurename}{Fig.}
\newcommand{\figref}[1]{\figurename~\ref{#1}}
\graphicspath{{./Figures/}} 
\ifx\diagnoselabel\undefined
	\relax
\else
	\makeatletter
	 \def\@testdef #1#2#3{%
		 \def\reserved@a{#3}\expandafter \ifx \csname #1@#2\endcsname
		\reserved@a  \else
	 \typeout{^^Jlabel #2 changed:^^J%
	 \meaning\reserved@a^^J%
	 \expandafter\meaning\csname #1@#2\endcsname^^J}%
	 \@tempswatrue \fi}
	\makeatother
\fi

\usepackage{graphicx}
\usepackage{tikz}
\usepackage{color}
\usepackage{pgfplots}
\usepackage{cite}
\pgfplotsset{compat=1.5}

\providecommand{\U}[1]{\protect\rule{.1in}{.1in}}

\def\ui{{\mathrm{i}}}

\begin{document}
	
\title{Folded Graph Signals: Sensing with Unlimited Dynamic Range}
\author{Feng~Ji, Pratibha, and Wee~Peng~Tay,~\IEEEmembership{Senior Member,~IEEE}%
	\thanks{This work was supported in part by the Singapore Ministry of Education Academic Research Fund Tier 2 grant MOE2018-T2-2-019 and by A*STAR under its RIE2020 Advanced Manufacturing and Engineering (AME) Industry Alignment Fund – Pre Positioning (IAF-PP) (Grant No. A19D6a0053).}%
	\thanks{The authors are with the School of Electrical and Electronic Engineering, Nanyang Technological University, 639798, Singapore (e-mail: jifeng@ntu.edu.sg, pratibha001@e.ntu.edu.sg, wptay@ntu.edu.sg).}
}
\maketitle
\begin{abstract}
Self-reset analog-to-digital converters (ADCs) are used to sample high dynamic range signals resulting in modulo-operation based folded signal samples. We consider the case where each vertex of a graph (e.g., sensors in a network) is equipped with a self-reset ADC and senses a time series. Graph sampling allows the graph time series to be represented by the signals at a subset of sampled vertices and time instances. We investigate the problem of recovering bandlimited continuous-time graph signals from folded signal samples.  We derive sufficient conditions to achieve successful recovery of the graph signal from the folded signal samples, which can be achieved via integer programming. To resolve the scalability issue of integer programming, we propose a sparse optimization recovery method for graph signals satisfying certain technical conditions. Such an approach requires a novel graph sampling scheme that selects vertices with small signal variation. The proposed algorithm exploits the inherent relationship among the graph vertices in both the vertex and time domains to recover the graph signal. Simulations and experiments on images validate the feasibility of our proposed approach. 
\end{abstract}


\section{Introduction}\label{sec:introduction}

Graph signal processing is an emerging field that studies multidimensional signals embedded in a graph representing the inherent relationship among the different components of the signals \cite{Shuman2013}. Graph signal processing has attracted increased attention as it allows us to capture complex correlations in many practical problems. It has been applied to various problems consisting of signal recovery, prediction, and anomaly detection \cite{Shuman2013, San13, San14, Gad14, Moura2015, Don16, Egi17, Ort18, JiTay:J19}. Recently, much work has been devoted to graph sampling, which studies the recovery of entire graph signal using observations at only some of the graph vertices \cite{Chen2015,Tsit2016,Anis2016,Qiu2017,Romero2017,JiTay:C18,tanaka2018}. 

\begin{figure}[!htb]
	\centering
	\includegraphics[scale=0.4]{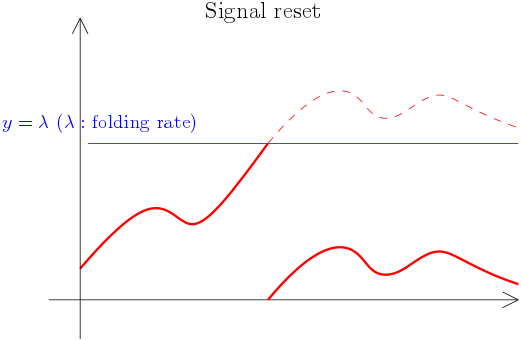}
	\caption{Signal reset with folding rate $\lambda$ results a folding signal depicted by the discontinuous solid red curve.}
	\label{fig:reset}
\end{figure}

Analog-to-digital converters (ADCs) are used to sample and digitize continuous signals. The Nyquist-Shannon rate is the minimum sampling rate that guarantees perfect recovery of a bandlimited signal from its signal samples \cite{Shannon1948}. To overcome the problem of signal clipping that occurs during sampling of high dynamic range signals, self-reset ADCs are used to sample modulo-operated signal values, which are called folded samples \cite{park2007, adc2009, chou1992,Ayush2017,Rudresh2018,Ayush2018}. Suppose $[0,\lambda]$ is the maximum amplitude range the ADC can capture. A signal $x$ and its folded version $0\leq p <\lambda$ (as illustrated in \figref{fig:reset}) are related by the following equation analogous to modulo arithmetic:
\begin{align} \label{eq:modulo}
x = \lambda z + p, 
\end{align}
where $z\in\bbZ$ is the largest integer not more than $x/\lambda$. We call $z$ the folding number.

We want to recover the original signal $x$ from the folded signal $p$ without knowledge of the folding number $z$. In this context, the authors in \cite{Ayush2017} introduced the concept of unlimited sampling, where a sufficient condition was derived for the recovery of bandlimited signals from the folded samples. They showed that using a sampling rate $e$ (Euler's number) times that of the Nyquist-Shannon rate is sufficient to recover a signal from its folded samples. An earlier related work is \cite{Chen14}. Both paper propose folded signal recovery based on the method of finite-difference and its higher order version. This approach encounters the difficulty when applied to graph signals, as there is no analogous operators on sampled subgraphs. The paper \cite{Gra19} proposed the notion of one-bit unlimited sampling to overcome the dynamic range limitations of a conventional one-bit quantizer. In \cite{Mus18}, the authors developed a generalized approximate message passing algorithm for the reconstruction of discrete-time sparse signals with noise. The reference \cite{Cuc18} studied the denoising problem of a smooth function on $[0,1]$ at discrete sampled points, while observing modulo $1$ samples. In \cite{Rudresh2018}, the authors considered folded signal recovery where the original signal are annihilated by certain wavelet transforms, in particular, polynomial signals. All the above mentioned works are taking signals from a $1$D channel. The authors in \cite{Ord18} investigated modulo sampling based hardware implementation and quantization. Signals are recovered from the folded samples together with complete signals from the past. The signals considered in this paper can be vector valued, however, no graph structured signals are considered. On the other hand, there are also considerable effects on the hardware side to build self-reset counter CMOS such as \cite{Cai09, Tak16}.

The signals observed in many applications can be modeled as graph signals, for example, photo and microscopic images \cite{Tak16}, and readings from sensor networks. This motivates us to study signal recovery from folded versions in the context of graph signals. Different from most of the other related works on folded signals described above that considers only a single continuous-time signal, we consider the problem of recovery of a \emph{continuous-time graph signal} in which each vertex of a graph is associated with a continuous-time signal \cite{JiTay:J19}. This allows us to perform discrete sampling in both the graph vertex and time domain, i.e., we employ a spatio-temporal sampling. Moreover, it presents an opportunity to exploit the additional correlation information captured in the graph signals to enhance the signal recovery methods. 

\begin{figure}[!htb]
 	\footnotesize
 	\centering
 	\begin{minipage}[b]{.5\linewidth}
 		\centering
 		\centerline{\includegraphics[scale=0.6]{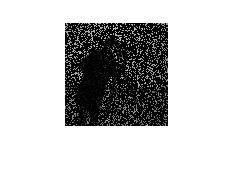}}
 		\centerline{(a)}
 	\end{minipage}%
 	\begin{minipage}[b]{.5\linewidth}
 		\centering
 		\centerline{\includegraphics[scale=0.6]{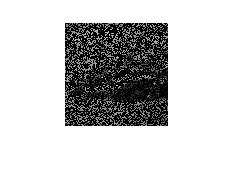}}
 		\centerline{(b)}
 	\end{minipage}
 	\caption{Examples of grayscale folded images. The recovered images are given in \figref{fig:7} of Section~\ref{sec:sr}.}
 	\label{fig:5}
 \end{figure}

If the signals do not have time components, we are restricted to graph signals in the traditional sense \cite{Shuman2013}. An important special case is imaging. Pixel values can be considered as signals on a grid graph. If the pixel values of an image are folded, we obtain a \emph{folded image} (e.g., \figref{fig:5}). The method proposed in the paper can be used to recover folded images as we shall illustrate with a few examples in Section~\ref{sec:sr}.

In this paper, we consider recovery of a graph signal bandlimited to the first $K$ eigenvectors of the graph shift operator in the vertex domain and to the baseband frequency band $[-B,B]$~Hz in the time domain. If the signal samples are \emph{unfolded}, the Nyquist-Shannon rate for perfect signal recovery is $2BK$~Hz, i.e., recovery can be achieved by sampling at a rate of $2B$~Hz at $K$ carefully chosen vertices of the graph \cite{JiTay:J19}. Our main contributions are summarized as follows:
\begin{enumerate}[(a)]
	\item We provide sufficient conditions to achieve graph signal recovery from a set of \emph{folded} samples with sampling rate $2B(K+1)$~Hz. In particular, we show that under a linear independence condition, only a single vertex signal in addition to the $K$ chosen vertices for unfolded recovery needs to be sampled at a rate of $2B$~Hz. Note that the sampling rate of $2B(K+1)$~Hz is lower than $2BKe$~Hz if we apply the finite-order difference approach of \cite{Ayush2017} to each vertex signal individually.
	\item To achieve practical recovery of a continuous-time graph signal from folded samples, we propose a sparse optimization procedure that is scalable. We introduce the novel concept of partition complexity for graph signals and show how to achieve minimal partition complexity through a greedy algorithm. We show that the degree of freedom in our sparse optimization, which we call $\lambda$-sparsity, is bounded by a partition complexity. This then allows us to formulate an effective sparse $L^1$ optimization problem.
	\item We apply our recovery method on various continuous-time graph signals and images, including a folded image from a mouse brain \cite{Tak16} and images from MIT-Adobe FiveK Dataset. We provide insights on how our proposed recovery method can be adapted in practice and discuss its performance.
\end{enumerate}
A preliminary version of this work was presented in \cite{JiPraTay:C19}. In this paper, we present rigorous proofs of all results and included further experiments and tests of our approach on folded network signals and images.

The rest of the paper is organized as follows. In Section~\ref{sec:fgs}, we present our system model and assumptions. We show that recovery of a graph signal bandlimited in both vertex and time domains can be achieved by sampling at an additional node using the Nyquist-Shannon rate in the time direction. In practice, the graph signal can be recovered by integer programming, which becomes intractable when the size of graph grows. To mitigate this, we introduce the concept of partition complexity in Section~\ref{sec:gs}, which then allows us to formulate a sparse optimization based algorithm for signal recovery in Section~\ref{sec:gsr}. The proposed recovery method leverages on the signal correlations in both the graph and time domain for efficient signal reconstruction. We present simulation and experiment results in Section~\ref{sec:sr} and conclude in Section~\ref{sec:c}. 

Notations. We use $\Real$ and $\bbZ$ to denote the set of real numbers and integers, respectively. Suppose $x : V \mapsto \Real$ is a mapping from a set $V$ to $\Real$. Then, for a subset $S\subset V$, we let $x(S) = (x(v))_{v \in S}$. For a matrix $W$, $W_S$ is the submatrix consisting of the rows of the matrix $W$ indexed by the set $S$. Thus, $W_u$ is the row $u$ of $W$.

\section{Folded graph signals} \label{sec:fgs}

In this section, we present our system model and assumptions. We also derive a sufficient condition under which perfect recovery from folded graph signal samples is achievable. Consider an undirected, simple graph $G=(V,E)$ with $V$ the set of vertices and $E$ the set of edges. Let $x = (x(v,t))_{v\in V, t\in\mathbb{R}}$ be a \emph{continuous-time graph signal}, where for each vertex $v\in V$, $x(v,\cdot) : \Real\mapsto\Real$ is a $L^2$ function \cite{JiTay:J19}. Component-wise, for each $t\in \mathbb{R}$, $x(\cdot,t) = (x(v,t))_{v\in V}$ is a graph signal; and for each $v\in V$, $x(v,\cdot) = (x(v,t))_{t\in \mathbb{R}}$ is a continuous-time signal at $v$. Note the $x$ is a generalized graph signal as defined by \cite{JiTay:J19}. We start with the following bandlimit assumptions. 

\begin{Assumption}\label{assumpt:BL1}
	For each time $t$, the graph signal $x(\cdot,t)$ is spanned by $W=\{w_1,\ldots, w_K\}$, $K$ eigenvectors of the associated eigenvalues of the Laplacian matrix $L$ of the graph $G$ \cite{Shuman2013}. 
\end{Assumption}

We use $W$ to denote the matrix with $w_i$ as the columns. At any time $t$, $x(\cdot,t)$ can be represented using the basis vector coefficients $b_t = (b_{k,t}) \in \mathbb{R}^K$ as 
\begin{align} \label{eq:xws}
x(\cdot, t) =  Wb_t = \sum_{1\leq k\leq K} b_{k,t}w_k.
\end{align}
As discussed in \cite{Chen2015,Tsit2016}, at each time $t$, we can sample the graph signal by recording the signal at a subset $S$ of $V$ to obtain the spatially sampled signal $x(S,t)$. The submatrix of $W$ formed by taking the rows indexed by $S$ is denoted by $W_S$. The sampled signal can be used to recover the full graph signal if the sampled nodes are chosen such that the corresponding basis matrix $W_S$ of size $K\times K$ is invertible.  

It is common to have $W$ consisting of eigenvectors corresponding to the $K$ smallest eigenvalues of $L$, for example, when one performs graph signal smoothing by using a low-pass filter.

\begin{Assumption}\label{assumpt:BL2}
At each node $v\in V$, the $L^2$ function $x(v,\cdot)$ is assumed to be bandlimited to $[-B,B]$~Hz.
\end{Assumption}

If we sample $x$ discretely with a subset $U\subset V\times \mathbb{R}$, a useful measure is the \emph{sampling rate} of $U$ defined as: 
\begin{align*}
\limsup_{a\to \infty}\frac{|(V\times [-a,a])\cap U|}{2a}.
\end{align*}
Using a sampling interval $T_0$, the discrete signal
\begin{align*} 
y=(y(v,n))_{v\in V, n\in\mathbb{Z}}=(x(v,nT_0))_{v\in V, n\in \mathbb{Z}} 
\end{align*}
is obtained. Applying the Nyquist-Shannon sampling rate, we may choose $T_0=1/(2B)$ for complete signal recovery of $x(v,\cdot)$ from the sampled discrete signal $y$. Using both spatial and temporal sampling, a sampling rate of $2BK$ over the entire graph guarantees a perfect recovery of the full signal. The concept of a $\calF$-transform for generalized graph signals was developed in \cite{JiTay:J19} to quantify the variation of the signal using the joint spectrum over the graph vertex and time domains. The $\calF$-transform of $x$ in \eqref{eq:xws} evaluated at the $k$-th graph eigenvalue and time domain frequency $f$, where $k\in\{1,\ldots,K\}$ and $f\in[-B,B]$, is given by
\begin{align*}
d_{k,f} = \frac{1}{2\pi}\int_\Real b_{k,t} e^{-\ui 2\pi t f}\ \ud t,
\end{align*}
which is also the Fourier transform of $b_{k,t}$ if we consider it to be a continuous-time signal. 

\begin{Assumption}\label{assumpt:BL3}
For each $k=1,\ldots,K$ and $f\in [-B,B]$, 
\begin{align}
|d_{k,f}| \leq a_{k,f}.
\end{align}
\end{Assumption}
Assumption~\ref{assumpt:BL3} is used in the quantitative analysis in Section~\ref{sec:lc}.  

We consider the scenario where the graph signal is not recorded perfectly by the ADC, e.g., the signal dynamic range exceeds the voltage range of the ADC. We use self-reset ADCs to sample the graph signal. A self-reset ADC captures the signal at each sampled node and yields a folded signal via a modulo operation. More specifically, the maximum amplitude that can be measured by the ADC, a positive real number $\lambda$, is called the \emph{folding rate}. Using the definition of a modulo operator described in \eqref{eq:modulo}, the graph signal $y(S,n) = (y(v,n))_{v\in S} \in \mathbb{R}^K$ can be expressed as follows:
\begin{equation} 
\label{eq:1}
y(S,n) = { D}_{\lambda}z(S,n) + p(S,n),
\end{equation}
where ${ D}_{\lambda}$ is a diagonal matrix with diagonal entries all equal to $\lambda$ and $0\leq p(v,n) < \lambda$ for all $v\in S$. We refer to $z(S,n)\in\mathbb{Z}^{K}$ as the vector of folding numbers and $p(S,n)\in [0,\lambda)^K$ as the vector of folded signals. 

For an invertible matrix $W_S$, the basis coefficients $b_{i,nT_0}$, and hence $y(\cdot,n)$, are uniquely determined by $y(S,n)$. However, if we only observe $p(S,n)$, it is in general not enough to recover $y(\cdot,n)$, and additional information is required. 

To state such information, we need one more notion: $k$ real numbers $a_1,\ldots, a_k$ are said to be \emph{linearly independent over $\mathbb{Z}$} if the only integer solution to the equation $\sum_{1\leq i\leq k} a_ix_i=0$ is $x_i=0$ for all $1\leq i\leq k$. It is common to have $k$-tuples linearly independent over $\mathbb{Z}$ by countability. 

\begin{Theorem}\label{thm:1}	
Consider graph signals that belong to the span of ${W}=\{{w}_1,\ldots, w_K\}$. Let $S\subset V$ be a subset of sampled nodes of size $K$ such that the submatrix $W_S$ formed by taking the rows of $W$ indexed by $S$ is invertible. Suppose that folded signals with folding rate $\lambda$ are observed at vertices in $S$. We have the following:
	\begin{enumerate}[(a)]		
		\item \label{it:lub} Let $u \in V\backslash S$ be an additional node such that the entries of $W_uW_S^{-1}$ are linearly independent over $\mathbb{Z}$. If the folding rate $\lambda'$ at $u$ is chosen according to a probability distribution absolutely continuous with respect to (w.r.t.) the Lebesgue measure, then with probability one, any graph signal $x : V \mapsto \Real$ can be recovered from the folded signals of $x$ at $V' = S\cup \{u\}$.
		
		\item If $u\in V\backslash S$ with folding rate $\lambda$ such that $W_uW_S^{-1}$ contains an irrational entry, then there are infinitely many $x$ with the same folded signals at $S$ and distinct folded signals at $u$.
	\end{enumerate}
\end{Theorem}

\begin{IEEEproof}
\begin{enumerate}[(a)]
\item Consider two graph signals ${x}=\sum_{1\leq i\leq K}{a}_{i}w_{i}$ and $\hat{x}=\sum_{1\leq i\leq K}\hat{a}_{i}w_{i}$, with the corresponding vectors of folded signals and folding numbers at the sample set $S$ given by $\{p,z\}$ and $\{\hat{p}, \hat{z}\}$, respectively. Suppose $p=\hat{p}$. We want to show that if $x$ and $\hat{x}$ have the same folded sample at $u$ with folding rate $\lambda'$ randomly chosen according to a distribution absolutely continuous w.r.t.\ the Lebesgue measure, then with probability one, $z = \hat{z}$ and hence $x=\hat{x}$.

Since $p=\hat{p}$, we have
\begin{align} \label{eq:s1l}
\sum_{1\leq i\leq K} (\hat{a}_{i}-a_{i})w_i = D_\lambda q, 
\end{align}
where $q= \hat{z}-z \in \bbZ^K$ is the difference in the folding numbers of the two graph signals $\hat{x}$ and $x$ at the sample set $S$. For a node $u \notin S$, write $W_u = (w_i(u))_{1\leq i\leq K}$. The following holds: 
\begin{equation} \label{eq:hxu}
\hat{x}(u)-x(u) = \sum_{1\leq i\leq K}(\hat{a}_{i}-a_{i})w_{i}(u). 
\end{equation}
Combining \eqref{eq:s1l} together with \eqref{eq:hxu}, and since $W_S$ is invertible, we have 
\begin{align} 
\hat{x}(u)-x(u) &= \lambda W_u { W}_S^{-1}q, \nn
& = \lambda \sum_{1\leq j\leq K} q_j (W_uW_S^{-1})_j \label{eq:hxx}, 
\end{align}
where $q_j$ and $(W_uW_S^{-1})_j $ are the $j$-th components of the vectors $q$ and $W_uW_S^{-1}$, respectively.

If $q_j, 1\leq j\leq K$ are not all $0$, then as the entries of $W_uW_S^{-1}$ are linearly independent over $\mathbb{Z}$, the expression $\lambda \sum_{1\leq j\leq K} q_j (W_uW_S^{-1})_j$ is non-zero. Moreover, if $q$ varies over $\mathbb{Z}^K\backslash \{0\}$, the collection $\Delta =\left\{ \lambda \sum_{1\leq j\leq K} q_j (W_uW_S^{-1})_j : q \in \bbZ^K\backslash\{0\}\right\}$ is a countable set in $\Real$. We have seen that $0\notin \Delta$. Hence, the set $\Lambda = \{\lambda' : \delta/\lambda' \in \mathbb{Z} \text{ for some } \delta\in \Delta\}$ is countable and has Lebesgue measure zero. Therefore, with probability one, a randomly chosen $\lambda'$ does not belong to $\Lambda$. Consequently, with such $\lambda'$, if $x$ and $\hat{x}$ has the same folded sample at $u$, then $q=0$.  

\item Without loss of generality, we assume that $r = (W_uW_S^{-1})_1$ is irrational. In \eqref{eq:hxx}, we let $q_j=0$ for $j>1$ and let $q_1$ vary over $\mathbb{Z}$. With such a choice, as $r$ is irrational, the folded values of $\hat{x}(u)-x(u) = \lambda q_1r$ are dense in the interval $[0,\lambda)$. In particular, there are infinitely many possibilities for $\hat{p}_u-p_u$. Therefore, we can observe the same folded signals at $S$ and different folded signals at $u$ for infinitely many $x$. 
\end{enumerate}	
	
\end{IEEEproof}

One should note that Theorem~\ref{thm:1}\ref{it:lub} does not rule out the possibility $\lambda' = \lambda$.  From the proof, we can choose $\lambda'=\lambda$ if $\sum_{1\leq j \leq K}q_j(W_S^{-1}W_u^T)_j$ is irrational for some $(q_j)_{1\leq j\leq K} \in \mathbb{Z}^K$. In Appendix~\ref{app:gre}, we demonstrate that the conditions in Theorem~\ref{thm:1} as well as the above condition for $\lambda'=\lambda$ are satisfied generally in certain random models. 

In practice, say a sensor network, the random folding rate at an additional node required in Theorem~\ref{thm:1}\ref{it:lub} can be realized by setting one of the sensor nodes to operate at a lower folding rate $\lambda'$ than its maximum voltage level $\lambda$, choosing it uniformly randomly in $(0,\lambda)$. We would like to point out that $\lambda'$ appearing in the theorem is independent of the signal $x$. Therefore, as long as $G$ and $W$ are fixed and the conditions of Theorem~\ref{thm:1}\ref{it:lub} are satisfied, we are able to recover the graph signal from folded samples. This leads to the following. 

\begin{Corollary}[Discrete sampling rate]
	Suppose $W,S,u$ and $\lambda'$ are the same as in Theorem~\ref{thm:1}\ref{it:lub}. There are discrete subsets $S_1,S_2 \subset V\times \mathbb{R}$ with sampling rate $2BK$ and $2B$ respectively such that the following holds: Consider any $x$ such that $x(\cdot,t)$ belongs to the span of columns of $W$ and $x(v,\cdot)$ is bandlimited by $B$ for any $t\in \mathbb{R}, v\in V$. The generalized graph signal $x$ can be recovered from the folded signals at $S_1$ with folding rate $\lambda$ together with folded samples at $S_2$ with folding rate $\lambda'$.
\end{Corollary}
\begin{IEEEproof}
Let $T'$ be a discrete subset of $\mathbb{R}$ sampled at the Nyquist-Shannon rate, i.e., with sampling rate $2B$. Take $S_1= S\times T'$ and $S_2=\{u\}\times T'$. For each $t \in S'$, we are able to recover the entire graph signal $x(\cdot,t)$ from the folded samples at $(S \cup \{u\})\times \{t\}$ by Theorem~\ref{thm:1}\ref{it:lub}. Consequently, by the Nyquist-Shannon sampling theorem, we can recover $x(v,\cdot)$ for each $v\in V$ and hence the entire signal $x$.
\end{IEEEproof}

Suppose we have a sample set $S$ of size $K$ with folding rate $\lambda$ and another sample set $S'$ of size $K'\geq 1$ with folding rate $\lambda'$ such that $W_S$ is invertible. In addition, we assume that for some $u \in S'$, the entries of $W_uW_S^{-1}$ are linearly independent over $\mathbb{Z}$. From Theorem~\ref{thm:1}\ref{it:lub}, the folded signals at $S\cup S'$ allow us to recover the generalized graph signal.  Specifically, at time $nT$, let the folded signals and folding numbers at $S$ and $S'$ be given by pairs of vectors $\{p(S,n),z(S,n)\}$ and $\{p(S',n),z(S',n)\}$, respectively. Using \eqref{eq:1} and noting that $y(S',n) = W_{S'}W_S^{-1}y(S,n)$, we have
\begin{align}\label{eq:lme}
W_{S'}W_S^{-1}p(S,n) - p(S',n) &= -W_{S'}W_S^{-1}D_{\lambda}z(S,n)+D_{\lambda'}z(S',n).
\end{align}
Recoverability of the signal $x$ is equivalent to \eqref{eq:lme} having a unique integer solution. We can thus apply integer programming to recover $z(S,n)$ and $z(S',n)$. This makes our approach local in nature. This means that in each stage of the recovery scheme, we only need to consider finitely many sampled nodes, concentrated in a short time interval. It can also be applied to snapshot graph signal recovery from observations of folded signals at appropriately chosen sample nodes. Moreover, it can also locally recover signals that fail to fulfill global bandlimited or $L^2$ assumptions in the time direction.

Regarding sampling rate, a direct application of the unlimited sampling algorithm proposed in \cite{Ayush2017} for the signal at each vertex of the graph yields an overall sampling rate of  $2BKe$. In our proposed approach, by optimally exploiting the graph correlation information, a sampling rate of $2B(K+1)$ over the entire graph is achievable by choosing $K'=1$. This rate is significantly smaller than $2BKe$ for large $K$ and $B$. However, integer programming is intractable for large $K$. In the following, we propose an alternative sparse recovery method that is however suboptimal. We utilize both the spatial and temporal correlations to recover the signal. The main idea is to choose a sample set of vertices $V' = S \cup S'$ so that it can be further partitioned into subsets on which the graph signal do not vary much at each time instant. Each of these subsets of vertices then share ``approximately'' the same folding number at each time instant. In the following section, we introduce the concept of \emph{partition complexity} and show how it is related to the degrees of freedom in the graph signal. 

\section{Graph sampling}\label{sec:gs}

Recall that \emph{sampling} on $G=(V,E)$ refers to selecting a subset of nodes $V'\subset V$, so that certain requirements are met. The signal recovery problem from folded signals is equivalent to the recovery of the folding numbers. In this paper, we want to sample nodes whose signals tend to have small differences. This is because, at such nodes, the folding numbers have greater chance to be similar. 

\subsection{Partition complexity} \label{sec:pcs}

\begin{figure}[!htb]
	\centering
	\includegraphics[scale=0.5]{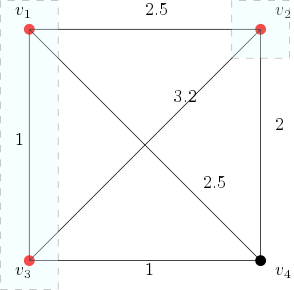}
	\caption{Let $V=\{v_1,v_2,v_3,v_4\}$ with $4$ nodes and $V'=\{v_1,v_2,v_3\}$. The values of the function $\phi$ are labeled along the edges. If $r = 2$, the complexity $c_{\phi,r}(V') = 2$. In this case, we apply the partition $V' = \{v_1,v_3\}\cup \{v_2\}$ with centers $v_1$ and $v_2$ respectively. On the other hand, if $r=3$, then $c_{\phi,r}(V')=1$. The partition $V'=V'$ is trivial, and the center must be chosen as $v_1$. Moreover, in the case $r=2$, we have $\min_{|V'|\geq 3} c_{\phi,r}(V')=1$. We can choose $V'=\{v_1,v_3,v_4\}$ with center $v_3$.}
	\label{fig:1}
\end{figure}

We need a way to quantify the difference between the signals at two different vertices. To do this, we utilize a symmetric non-negative ``distance'' function that will be carefully designed in the next Section~\ref{sec:lc} to account for the graph signal's bandlimitedness in both the graph vertex and time domains. We then wish to partition a set of sampled vertices into subsets of vertices that are close to each other in terms of this ``distance'' function. We start with a basic definition. See \figref{fig:1} for a simple illustration.

\begin{Definition}\label{def:phi}
Suppose $\phi: V\times V \to \mathbb{R}_{\geq 0}$ is a symmetric non-negative function on pairs of vertices of $G$. Let $r>0$ be a real number and $V'\subset V$ be a subset of nodes. A $(\phi,r)$-admissible partition of $V'$ is a decomposition of $V'$ into a union of subsets $V' = \bigcup_{1\leq i\leq k}V_i$ such that for each \emph{component} $V_i$, there is a \emph{center} $v_i\in V_i$ and $\phi(v_i,v)< r$ for all $v\in V_i$. The smallest $k$ such that $V'$ has an admissible partition into $k$-components is called the \emph{$(\phi,r)$-complexity} of $V'$, denoted by $c_{\phi,r}(V')$. 
\end{Definition}

For fixed $\phi$ and $r$, as we will see in Section~\ref{sec:lc}, it is desirable to find a subset of nodes $V'$ with size bounded below by an integer $s\leq |V|$ such that the $(\phi,r)$-complexity of $V'$ is as small as possible. This is re-cast as the following optimization problem: 
\begin{align} \label{eq:mvc}
\min_{|V'|\geq s} c_{\phi,r}(V').
\end{align}
To solve Problem \eqref{eq:mvc}, we can consider it from a different angle. For each $v \in V$, the $\phi$-ball of radius $r$ at $v$ is defined as the set $B_{\phi,r}(v)= \{v'\in V : \phi(v,v')< r\}$. For each $(\phi,r)$-admissible partition $V' = \cup_{1\leq i\leq k}V_i$, we have $V_i\subset B_{\phi,r}(v_i)$, where $v_i$ is a center. Therefore, Problem~(\ref{eq:mvc}) is equivalent to the following optimization problem:
\begin{align} \label{eq:mcg}
\min_{\substack{|\cup_{1\leq i\leq k} B_{\phi,r}(v_i)|\geq s,\\ v_1,\ldots, v_k\in V}} k.
\end{align}

\begin{Definition}
	For a fixed integer $0<k\leq |V|$ and a subset $S$ of $V$ of size $k$, we use $B_{\phi,r}(S)$ to denote the union $\cup_{s\in S}B_{\phi,r}(s),$ and call this the $\phi$-ball centered around $S$.
\end{Definition}

To solve Problem~(\ref{eq:mcg}), we may estimate $\max_{|S|=k}|B_{\phi,r}(S)|$, starting from $k=1$. Regarding the size of $B_{\phi,r}(S)$, we have the following observation. 

\begin{Lemma} \label{lem:tfn}
	The function $S\mapsto |B_{\phi,r}(S)|, S\subset V$ is a submodular set function.
\end{Lemma}

Recall that there are several equivalent characterization for submodularity and we describe one such characterization here. Let $\Omega$ be a set. A function $F: 2^{\Omega} \to \mathbb{R}$ on subsets of $\Omega$ is \emph{submodular} if and only if the following holds: for any $X\subset Y\subset \Omega$ and $x \in \Omega \backslash Y$, we have:
\begin{align}\label{submod}
F(X\cup \{x\}) - F(X) \geq F(Y\cup \{x\}) - F(Y).
\end{align}
We now proceed to the proof of Lemma~\ref{lem:tfn} with \eqref{submod} as the definition of submodularity.

\begin{IEEEproof}
Given $S_1\subset S_2\subset V$ and $v\in V$, we have the obvious inclusion 
\begin{align*}
B_{\phi,r}(v)\backslash B_{\phi,r}(S_2) \subset B_{\phi,r}(v)\backslash B_{\phi,r}(S_1).
\end{align*} 
Therefore, the function $S\mapsto |B_{\phi,r}(S)|$ is submodular following the above definition. 
\end{IEEEproof}

Maximizing a submodular function with size constraint is usually intractable. However, the \emph{greedy algorithm} gives a reasonable approximation \cite{Fei07}. Therefore, we propose the following algorithm for Problem~(\ref{eq:mcg}): we initialize by choosing $S=\{v\}$ such that $|B_{\phi,r}(v)|$ is maximized. In each subsequent iteration, we add a new node $v$ to $S$ such that $|B_{\phi,r}(S\cup \{v\})|$ is maximized. The procedure terminates when $|B_{\phi,r}(S)|\geq s$. 

\subsection{\texorpdfstring{$\lambda$}{lambda}-sparsity} \label{sec:lc}

We start with the following notion of $\lambda$-sparsity, which quantifies the degrees of freedom the folding numbers of a signal can have.

\begin{Definition}
	Given a vector $x = (x_i)_{1\leq i\leq n}$ written as $x = { D}_{\lambda} z+ p$ with $z \in \mathbb{Z}^n$ and $0\leq p_i<\lambda, 1\leq i\leq n$, the \emph{$\lambda$-sparsity} of $x$ is defined as the smallest size of a subset $I \subset \{1,\ldots, n\}$ such that $z = (z_1,\ldots,z_n)$ is uniquely determined by $\{z_i : i\in I\}$.
\end{Definition}

Intuitively, given a signal $x$, we would like to use $\lambda$-sparsity to count the amount of ``different" folding numbers when $x$ is folded w.r.t.\ $\lambda$. We now apply the results in Section~\ref{sec:pcs} by introducing an explicit non-negative symmetric function $\phi: V\times V \to \mathbb{R}_{\geq 0}$ as follows: 
\begin{align} \label{eq:phi}
\phi(u,v) &= \sqrt{2}\sum_{1\leq i\leq K} \int_{-B}^B  a_{i,f} \sqrt{1-\cos \frac{\pi f}{B}}\ \ud f \cdot |w_i(u)-w_i(v)|,
\end{align}
with $B$ as in Assumption~\ref{assumpt:BL2}, and $a_{i,f}$ as in Assumption~\ref{assumpt:BL3}.

Consider a bandlimited continuous-time graph signal $x$. Suppose we choose a discrete subset of samples $V'\times T$ in $V\times \mathbb{R}$, where $T=\{t_i : i\in \mathbb{Z}\}$ are time instances uniformly spaced by $1/(2B)$, i.e., at the Nyquist-Shannon rate. Let $x(V',\cdot)$ be the restriction of the signal $x$ to $V'$. To elucidate the reason behind our choice of $\phi$, we have the following estimate on the $\lambda$-sparsity regarding $x(V',\cdot)$.

\begin{Theorem} \label{thm:tls}
For any $i\in\bbZ$, the $\lambda$-sparsity of the difference $x(V', t_i)-x(V', t_{i+1})$ is upper bounded by the $(\phi, \lambda/2)$-complexity $c_{\phi,\lambda/2}(V')$ of $V'$. 
\end{Theorem}
\begin{IEEEproof}
We first remark that given $y_1=\lambda z_1 + p_1$ where $p_1$ is the folded signal of $y_1$, and given the folded signal $p_2$ of $y_2=\lambda z_2+p_2 \in (y_1-\lambda/2,y_1+\lambda/2)$, then the folding number $z_2$ of $y_2$ is uniquely determined (see \figref{fig:injective}): if $p_2-p_1\geq \lambda/2$, then the $z_2=z_1-1$. If $p_1-p_2\geq \lambda/2$, then $z_2=z_1+1$. Otherwise, $z_2=z_1$. 
	
\begin{figure}[!htb]
\centering
\includegraphics[scale=1]{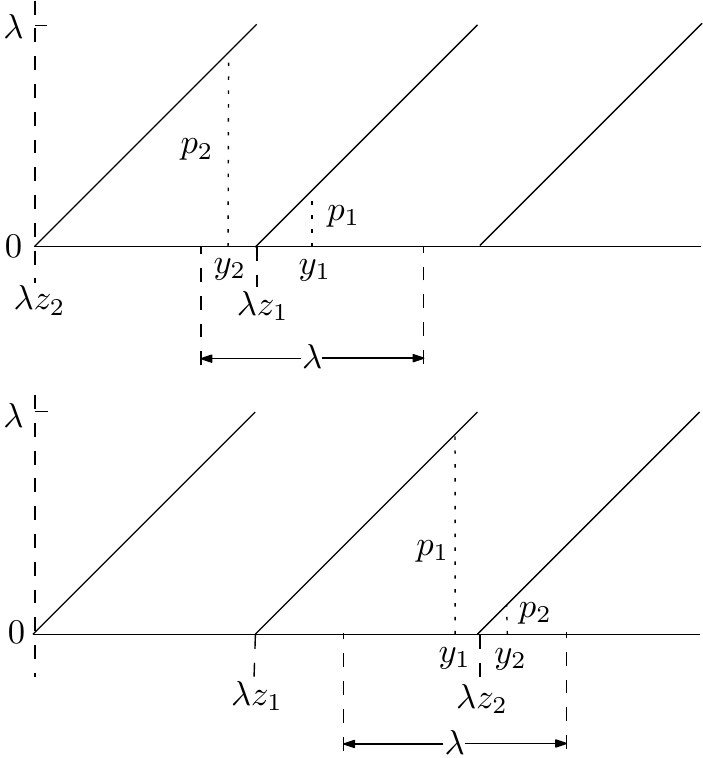}
\caption{The modulo operation is injective over an interval of length $\lambda$. Therefore, if $|y_2-y_1|<\lambda/2$, then the folding number of $y_2$ is uniquely determined by the folding number of $y_1$.}
\label{fig:injective}
\end{figure}	
	
With Assumption~\ref{assumpt:BL2}, we can express the continuous-time graph signal $x$ as
\begin{align*}
x(v,t) = \sum_{i=1}^K \int_{-B}^B d_{i,f}e^{\ui 2\pi t f}\ \ud f \cdot {w}_i(v),
\end{align*}
for a family of coefficients $d_{i,f}, 1\leq i\leq K, f \in [-B,B]$ such that $|d_{i,f}| \leq a_{i,f}$.

If $t_{i+1}-t_i = 1/(2B)$ and $u, v\in V'$, then 
\begin{align}
&|(x(u,t_i)- x(u,t_{i+1}))-(x(v,t_{i})- x(v,t_{i+1}))| \nn 
& = \abs*{\sum_{i=1}^K \int_{-B}^Bd_{i,f}(e^{2\pi \ui t_{i+1} f}-e^{2\pi \ui t_i f})\ \ud f \cdot (w_i(u)-w_i(v))} \nn
& \leq \sum_{i=1}^K\int_{-B}^Ba_{i,f}\cdot \sqrt{2}\sqrt{1-\cos (\frac{\pi f}{B})}\ \ud f \cdot |w_i(u)-w_i(v)| \nn
& = \phi(u,v), \label{phibound}
\end{align}
where the inequality follows from the assumption that $|d_{i,f}|\leq a_{i,f}$ and the equality
\begin{align*}
\abs*{e^{\ui2\pi(t+1/(2B)) f}-e^{\ui 2\pi t f}} = \sqrt{2\parens*{1-\cos \frac{\pi f}{B}}}, \text{ for } t\in \mathbb{R}.
\end{align*}

For convenience, let $y = x(V', t_i)-x(V', t_{i+1})$. Suppose 
\begin{align*}
V' = \bigcup_{1\leq i\leq c_{\phi,\lambda/2}(V')}V_i
\end{align*}
is an admissible partition with centers $v_i\in V_i$. Then for each $v\in V_i$, we have $\phi(v_i,v)<\lambda/2$ and from \eqref{phibound}, $|y(v_i)-y(v)|<\lambda/2$. As a consequence, the folding number of $y(v)$ is determined by the folding number of $y(v_i)$. Consequently, the $\lambda$-sparsity of $y$ is at most $c_{\phi,\lambda/2}(V')$.
\end{IEEEproof}

In applications, we may relax $\lambda/2$ by allowing a perturbation $\lambda/2+\epsilon$ with some suitable choice of $\epsilon>0$. The consequence is that we may have a smaller partition complexity, at the cost of losing $c_{\phi, \lambda/2+\epsilon}$ being the theoretical upper bound of $\lambda$-sparsity. This can be useful if the $a_{i,f}$'s in $\phi$ are not tight. We explore the effect of adding $\epsilon$ in Section~\ref{sec:sr}.

\section{Graph signal reconstruction} \label{sec:gsr}

In this section, we make use of the concepts introduced in the previous section to develop an algorithm that recovers a generalized graph signal using the folded samples recorded by self-reset ADCs. This is equivalent to recovering the folding numbers. 

Suppose we have samples $S$ of size $K$ with folding rate $\lambda$ and $S'$ of size $K'$ with folding rate $\lambda'$ such that $W_S$ is invertible. In addition, we assume that for some $u \in S'$, the entries of $W_uW_S^{-1}$ are linearly independent over $\mathbb{Z}$. From Theorem~\ref{thm:1}, these properties guarantee theoretical perfect recovery. See also Appendix~\ref{app:gre} for sufficient conditions under which these properties hold. At each time $nT$, where $T=1/(2B)$ and $n\in\bbZ$, let the folded signals and folding numbers at $S$ and $S'$ be given by pairs of vectors $\{p(S,n),z(S,n)\}$ and $\{p(S',n),z(S',n)\}$, respectively. 

\begin{Lemma} \label{lem:itc}
If the continuous-time graph signal $x$ is in $L^2(V\times \mathbb{R})$, then $z=(z(v,t))_{v\in V, t\in\Real}$ where $z(v,t)$ is the folding number of $x(v,t)$, is compactly supported.  
\end{Lemma}
\begin{IEEEproof}
As $V$ is finite, it suffices to show that $z(v,\cdot)$ is compactly supported for each $v \in V$. As $x \in L^2(V\times \mathbb{R})$, $x(v,\cdot)$ belongs to $L^2(\mathbb{R})$.	By Assumption~\ref{assumpt:BL2} and the Whittaker-Shannon interpolation formula \cite{Sha49}, 
\begin{align*}
x(v,t) = \sum_{n\in \mathbb{Z}} x(v,nT)\sinc\parens*{\frac{t}{T}-n},
\end{align*}
for $T=1/(2B)$. As integer translates of the $\sinc$ function are pairwise orthogonal with the same $L^2$-norm, the sum $\sum_{n\in \mathbb{Z}} |x(v,nT)|^2$ is finite. In particular, $|x(v,nT)| \to 0$ as $|n|\to \infty$ and $|x(v,nT)|$ is bounded, say by $b$. For $N_0\in \mathbb{Z}$, we have
\begin{align*}
|x(v,t)| &= \parens*{\sum_{n\in \mathbb{Z}} x(v,nT)\sinc(t/T-n)} \\ 
& \leq \sum_{n<N_0}|x(v,nT)|\cdot |\sinc(t/T-n)| + \sum_{n\geq N_0}|x(v,nT)|\cdot |\sinc(t/T-n)| \\
& \leq b \sum_{n<N_0}|\sinc(t/T-n)| + \sup_{n\geq N_0}\{|x(v,nT)|\}\sum_{n\in \mathbb{Z}} |\sinc(t/T-n)|.
\end{align*}
We can always choose $N_0$ such that $\sup_{n\geq N_0}\{|x(v,nT)|\}$ is arbitrarily small. At the same time, for any $|t|$ large enough, $\sum_{n<N_0}|\sinc(t/T-n)|$ can be made arbitrarily small. Therefore, $|x(v,t)| \to 0$ as $|t| \to \infty$. Thus, as $|t| \to \infty,$ $x(v,t)$ does not fold, and $z$ must be compactly supported.
\end{IEEEproof}

Recall from \eqref{eq:1} that $y(\cdot,n)$, $n\in\bbZ$, are samples of $x$ at discrete time instances. For square integrable bandlimited $x$, by Lemma~\ref{lem:itc}, the discrete samples $y(\cdot,n)$ are unfolded for large $|n|$. Hence, it suffices to recover the folding number for each $y(\cdot,n+1)-y(\cdot,n)$. In the case of clipping, i.e., we do not observe $y(\cdot,n)$ when $|n|$ is large, our method recovers the folding numbers up to an additive constant as in \cite{Ayush2017}. 

Fix an $n$ and let $\bar{y}(\cdot) = y(\cdot,n+1)-y(\cdot,n)$. Denote the folding numbers and folded signals of $\bar{y}(\cdot)$ by $\bar{z}(\cdot)$ and $\bar{p}(\cdot)$, i.e., $\bar{y}(\cdot) = D_{\lambda}\bar{z}(\cdot)+\bar{p}(\cdot)$. From \eqref{eq:lme}, we obtain
\begin{equation}\label{eq:wbd}
W_{S'}W_S^{-1}\bar{p}(S)  -\bar{p}(S')= D_{\lambda}\bar{z}(S')-W_{S'}W_S^{-1}D_{\lambda}\bar{z}(S).
\end{equation}
Our signal recover steps are as follows:
\begin{enumerate}[Step 1:]
	\item For a fixed $K'\geq 1$, we proceed by using the greedy algorithm to solve Problem~\eqref{eq:mcg} in Section~\ref{sec:pcs} to obtain a sample set of vertices $V'$ of size $K+K'$ such that $V'$ has (approximately) minimal $(\phi,\lambda/2)$-complexity. Decompose $V' = S\cup S'$ such that $|S|=K$ and $W_S$ is invertible. Write $V' = \cup_{1\leq i\leq c}V_i$ as a disjoint partition where $V_i$ has center $c_i$. 
	
	\item We replace the unknowns $\{\bar{z}(S), \bar{z}(S')\}$ in \eqref{eq:wbd} by variables $\zeta=(\zeta(v))_{v\in V'}$ as follows (cf. proof of Theorem~\ref{thm:tls}): 
\begin{enumerate}[(a)]
	\item If $v = c_i$ for some $i=1,\ldots,c$, make the substitution $\bar{z}(c_i)=\zeta(c_i)$.
	
	\item \label{it:stn} Suppose that $v\in V_i$. If $\bar{p}(v)-\bar{p}(c_i)\geq \lambda/2$, write $\bar{z}(v) = \zeta(c_i)+\zeta(v)-1$. If $\bar{p}(c_i)-\bar{p}(v)\geq \lambda/2$, write $\bar{z}(v) = \zeta(c_i)+\zeta(v)+1$. For the remaining cases, write $\bar{z}(v) = \zeta(c_i)+\zeta(v)$.   
\end{enumerate}

\begin{figure}[!htb]
	\centering
	\includegraphics[scale=0.5]{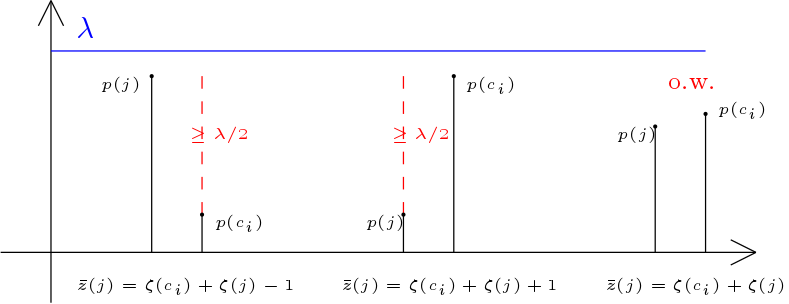}
	\caption{Illustration of the substitution of $\bar{z}$ by $\zeta$ in Step 2.} \label{fig:10}
\end{figure}

The above substitution is illustrated by \figref{fig:10}. In \ref{it:stn}, if $v\in V_i$ and the signal at vertices $c_i$ and $v$ differ less than $\lambda/2$, then $\zeta(v)=0$. Hence, if the above holds for most such pairs $c_i$ and $v$, we have a sparse solution in the variables $\zeta(v)$. 

\item We obtain a new set of equations on the $K+K'$ variables $\zeta = (\zeta(v))_{v\in V'}$ expressed as 
\begin{align} \label{eq:m}
M\zeta = g 
\end{align}
with $M$ being a $(K+K')\times K'$ matrix. As we observed earlier, $\zeta(v)=0$ unless $v = c_i$ for some center. In this case, $\zeta$ in (\ref{eq:m}) can be solved exactly if $\rank(M)\geq c_{\phi,\lambda/2}(V')$ by Theorem~\ref{thm:tls}. However, if the bound $a_{i,f}$ in Assumption~\ref{assumpt:BL3} is not tight, neither is the bound by the function $\phi$ in \eqref{phibound}. In this case, the $(\phi,\lambda/2)$-complexity of any $V' \subset V$ might be large. We may instead sample $V'$ with small $(\phi,\lambda/2+\epsilon)$-complexity for $\epsilon>0$. We no longer expect \eqref{eq:m} can be solved directly. However, at the compensation of a smaller partition complexity $c_{\phi,\lambda/2+\epsilon}(V')$, we expect $\zeta$ is sparse. Therefore, we heuristically propose solving:
\begin{align} \label{eq:mm}
\begin{aligned}
\min\ & \norm{\zeta}_1, \\
\st\ & M\zeta=g.
\end{aligned} 
\end{align} 
Moreover, suppose that at vertex $v$ the folding number $\bar{z}(v)$ is observed and $v$ belongs to $V_i$. This gives us an additional equation taking one of the forms: $\bar{z}(v) = \zeta(c_i)+\zeta(v)-1$, $\bar{z}(v) = \zeta(c_i)+\zeta(v)+1$  or $\bar{z}(v) = \zeta(c_i)+\zeta(v)$, which can be added to the linear constraints of Problem~(\ref{eq:mm}).
\end{enumerate}

\section{Simulation and Experiment Results}\label{sec:sr}

In this section, we present simulation results to demonstrate the performance of our method on folded continuous-time graph signals. We then conduct experiments on folded images to validate that our approach can recover the original unfolded image. To the best of our knowledge, our work is the first to consider the recovery of folded graph signals. Therefore, there are no existing benchmarks that we can compare with.

\subsection{Folded continuous-time graph signals}
\begin{figure}[!htb]
	\centering
	\includegraphics[scale=1.2]{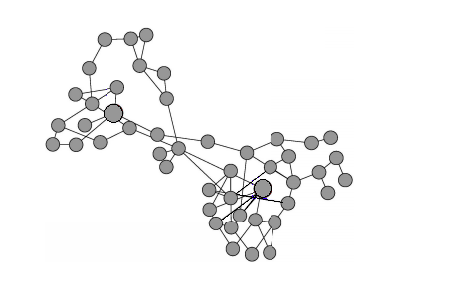}
	\caption{A part of the Arizona power plant network (modified from Fig.~8 of \cite{Luo2013}).} \label{fig:8}
\end{figure}

We consider $G$ being one of the following graphs: the complete graph with discrete random edge weights, the Arizona power plant network \cite{Ari12} as illustrated in \figref{fig:8}, and $2$-dimensional (2D) lattice of size $25\times 20$. A power plant may observe signals with high dynamic range such as temperature measurements \cite{Jet13}. On the other hand, a lattice can be used to model an image carrying HDR pixel values. For the parameters in Assumptions~\ref{assumpt:BL1}-\ref{assumpt:BL3}, we standardize the bandlimit $B$ to be $1$~Hz and set $a_{i,f}$ to be inversely propositional to $i$ and $f$. The choice of $K$ and $K'$ are summarized in Table~\ref{tab:1}.

\begin{table}[!htb]
	\caption{$K$ and $K'$ for different graphs.} \label{tab:1}
	\centering  
	\scalebox{1}{
		\begin{tabular}{|l|c|c|c|c|}  
			\hline
			{ Graph} & { size} & { $K$} & { $K'$}\\ 
			\hline\hline
			Complete graph & $120$ & $50$ & $20$  \\
			\hline
			Power plant & $47$ & $20$ & $10$  \\
			\hline
			$2$D lattice & $500$ & $80$ & $30$  \\
			\hline
	\end{tabular}}
	
\end{table}    

For each simulation instance, each coefficient of the signal $x$ w.r.t.\ the basis is chosen uniform randomly within the bounded set by the corresponding $a_{i,f}$. The folding rate $\lambda$ is chosen to ensure significant amount of non-zero foldings. In addition, we add white Gaussian noise to the observed folded signals with the signal-to-noise ratios (SNR) being $15$~dB, $20$~dB, and $40$~dB. As discussed in Section~\ref{sec:gsr}, we aim at recovering the difference signals. An example of a continuous-time graph signal and its folded version is shown in \figref{fig:gs_ex}. 

Due to noise, we apply the proposed recovery algorithm by solving the following modified version of Problem~(\ref{eq:mm}) to recover the difference signal: 
\begin{align} 
\min \norm{\zeta}_1 + \alpha \norm{M\zeta-g}_2^2, 
\end{align} 
where $\alpha$ is a chosen regularizing parameter.

\begin{figure}[!htbp]
	\footnotesize
	\centering
	\begin{minipage}[b]{.5\linewidth}
		\centering
		\centerline{\includegraphics[width=7.8cm, height=6.5cm]{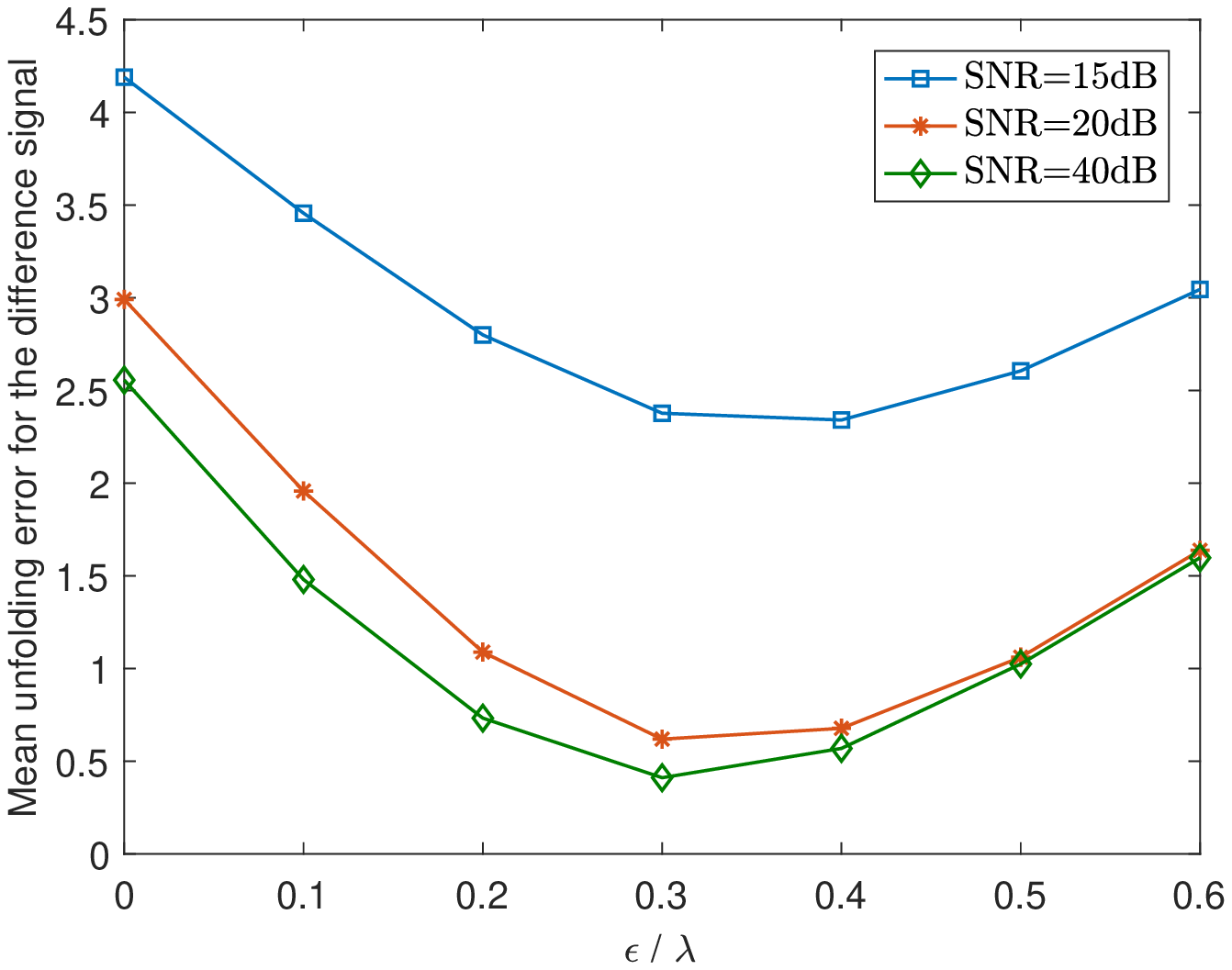}}
		\centerline{(a) Complete graph}
	\end{minipage}%
	\begin{minipage}[b]{.5\linewidth}
		\centering
		\centerline{\includegraphics[width=7.8cm, height=6.5cm]{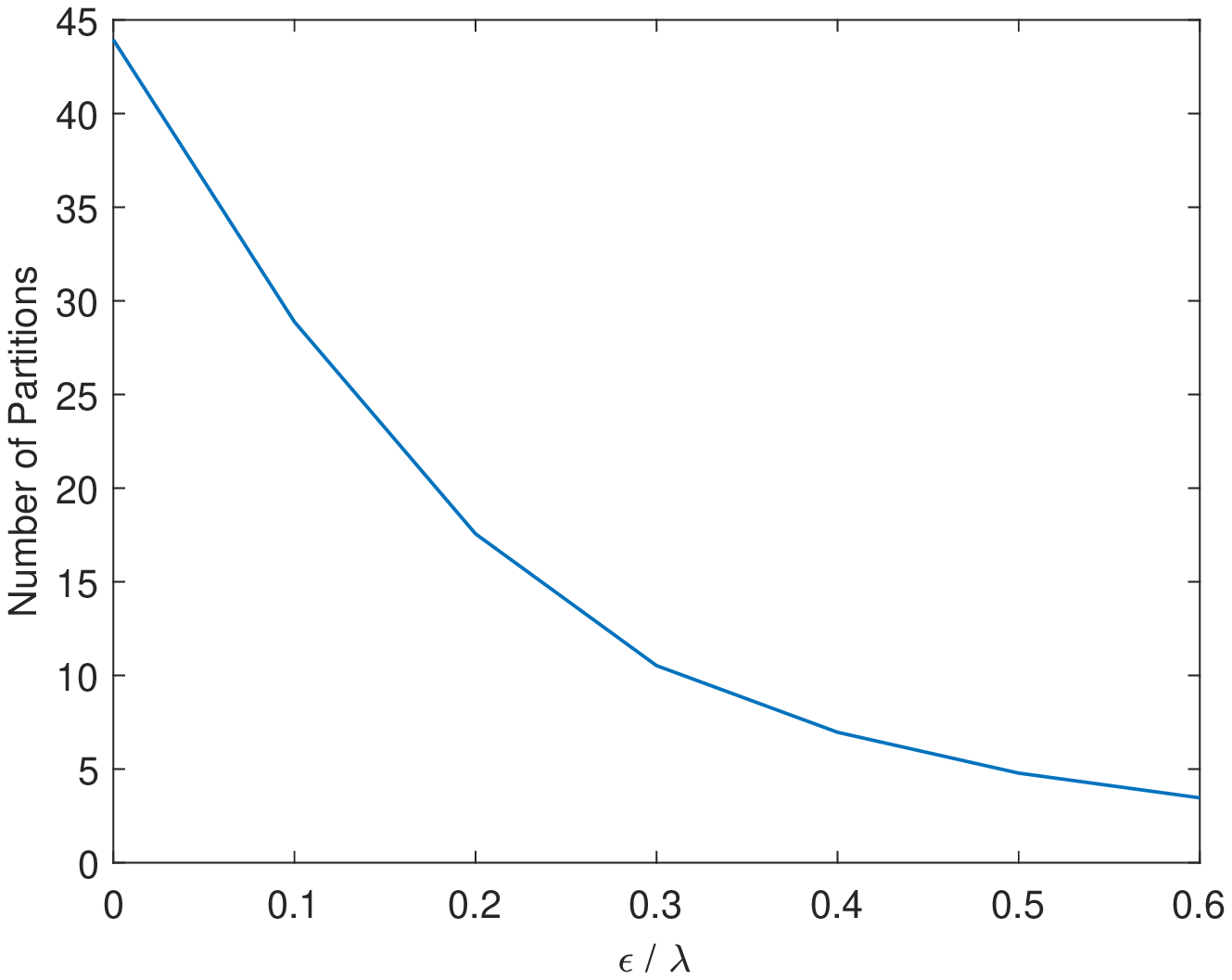}}
		\centerline{(b) Complete graph}
	\end{minipage}
	\begin{minipage}[b]{.5\linewidth}
		\centering
		\centerline{\includegraphics[width=7.8cm,height=6.5cm]{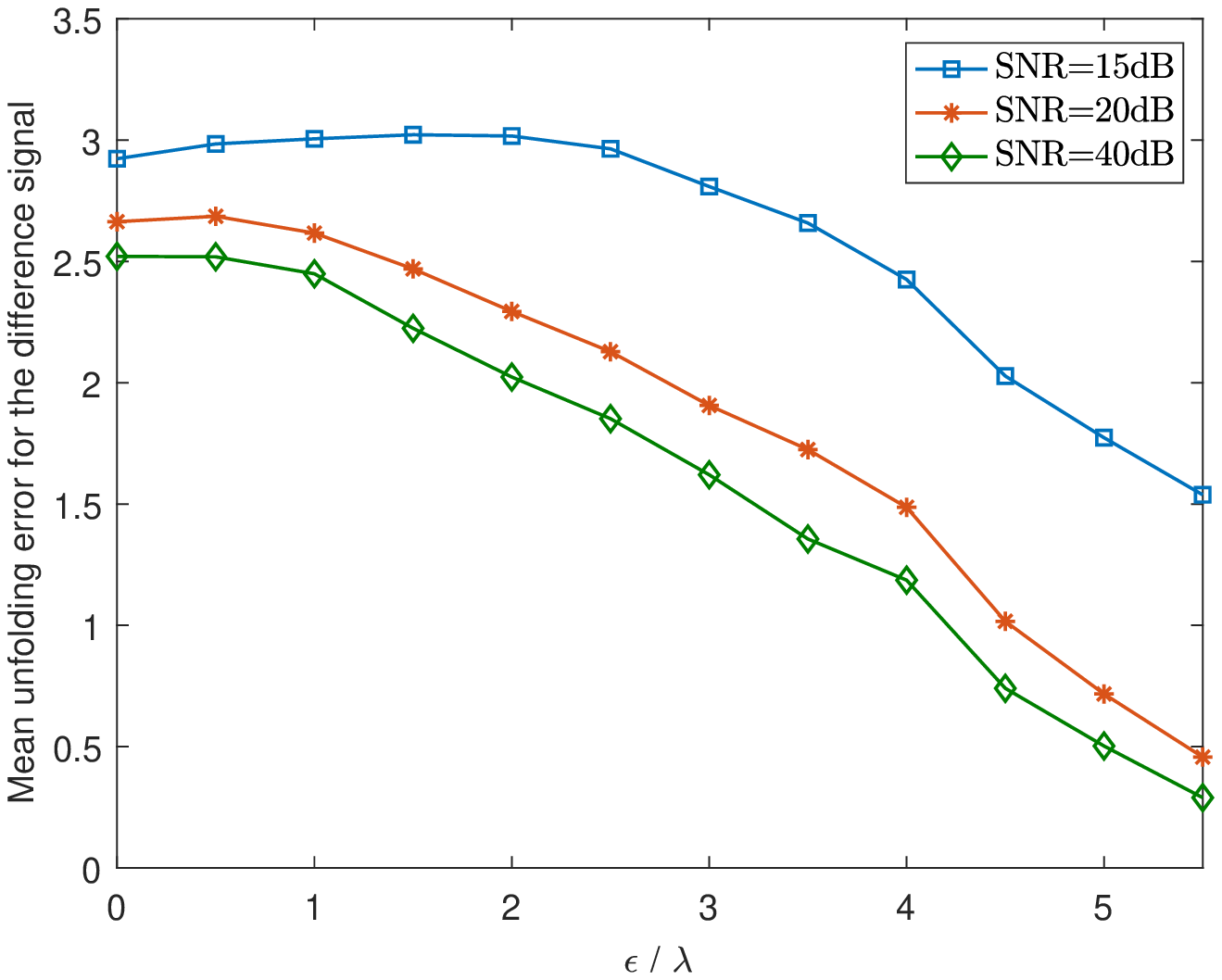}}
		\centerline{(c) Power plant}
	\end{minipage}%
	\begin{minipage}[b]{.5\linewidth}
		\centering
		\centerline{\includegraphics[width=7.8cm, height=6.5cm]{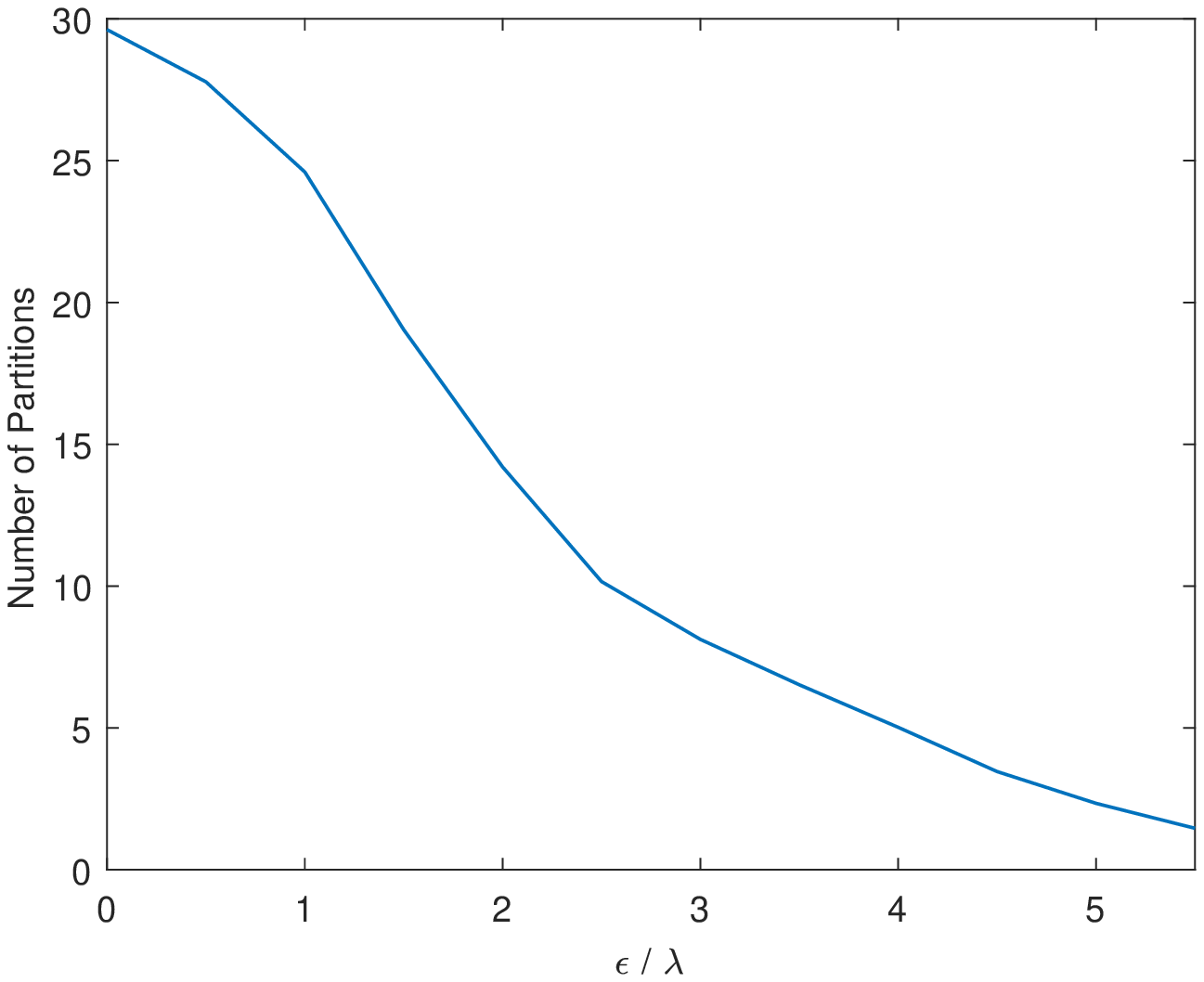}}
		\centerline{(d) Power plant}
	\end{minipage}
   	\begin{minipage}[b]{.5\linewidth}
   	\centering
   	\centerline{\includegraphics[width=7.8cm,height=6.5cm]{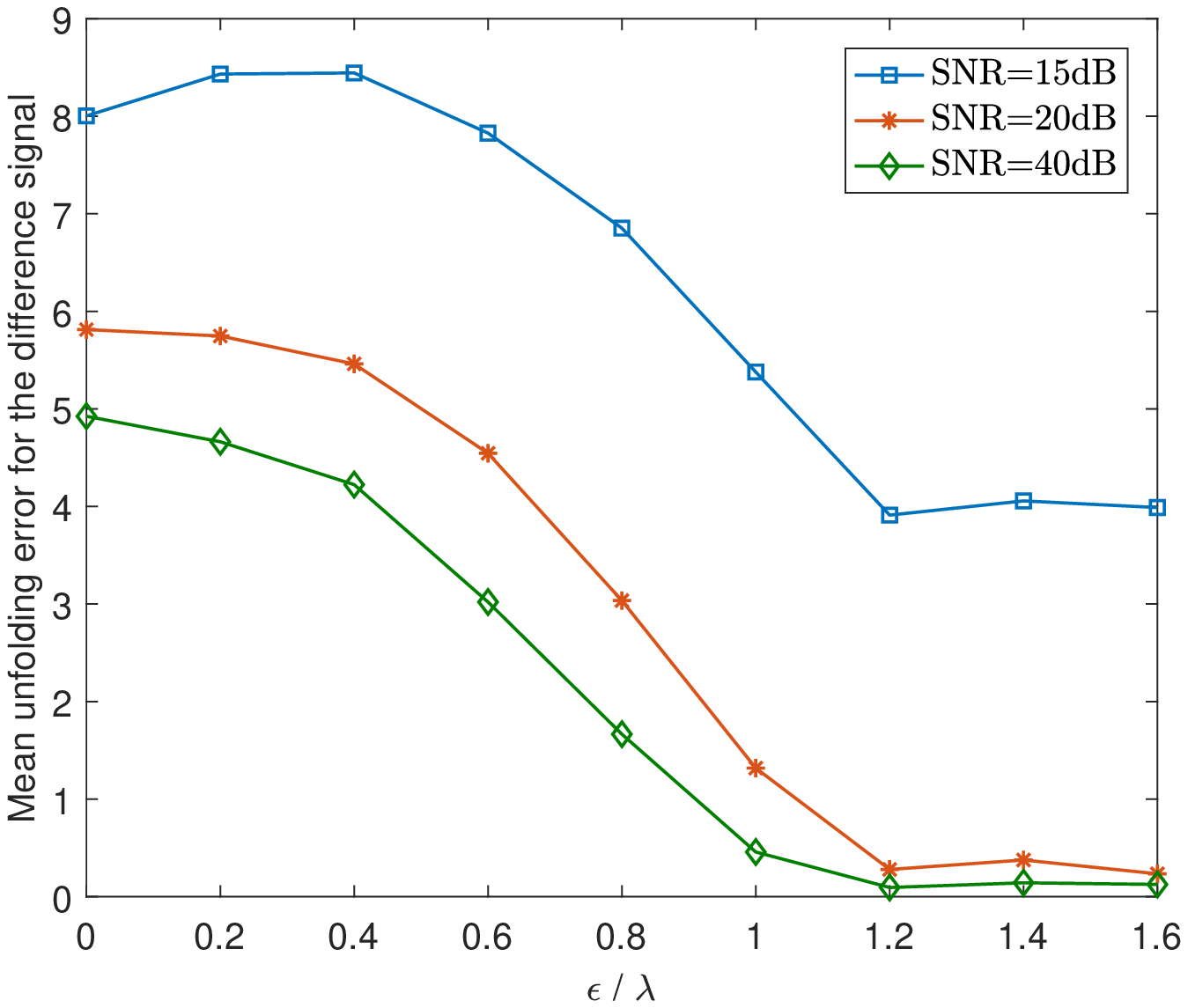}}
   	\centerline{(e) $2$D lattice}
   \end{minipage}%
   \begin{minipage}[b]{.5\linewidth}
   	\centering
   	\centerline{\includegraphics[width=7.8cm, height=6.5cm]{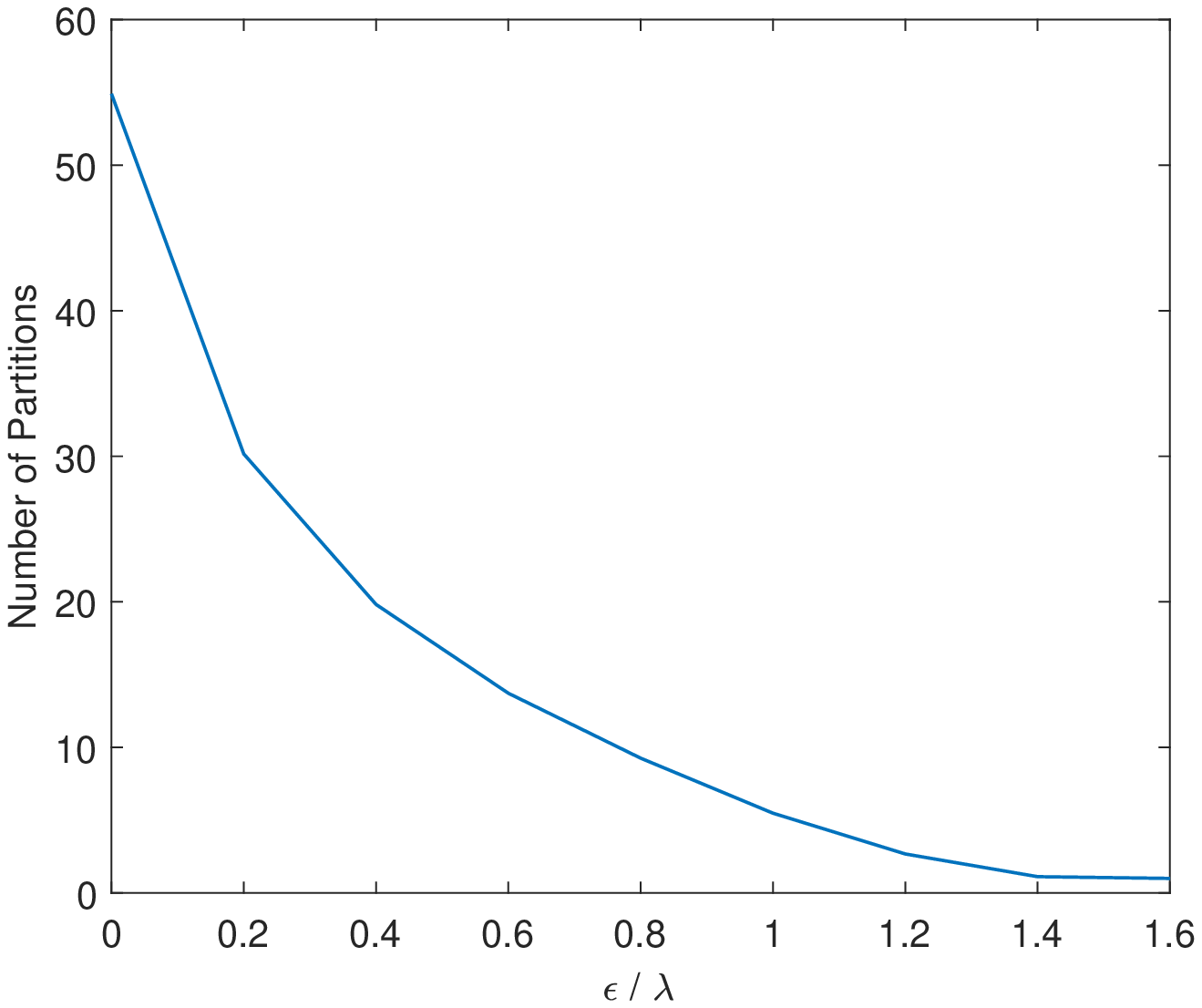}}
   	\centerline{(f) $2$D lattice}
   \end{minipage}
	\caption{The average total recovery error in folding numbers of the difference signals and the size of $(\phi, \lambda/2+\epsilon)$ admissible partition of $V'$ against $\epsilon/\lambda$. }
	\label{fig:2}
\end{figure} 

We choose the set of sample nodes $V'$ with small $(\phi, \lambda/2+\epsilon)$-complexity by the greedy algorithm. For different values of $\epsilon$, we plot in \figref{fig:2} the average total recovery error in the folding numbers and the size of $(\phi, \lambda/2+\epsilon)$ admissible partition of $V'$ against $\epsilon/\lambda$. Different curves correspond to different noise levels.

From the plots, we first notice that the performance gets better with less noise as expected. The errors are small if SNR = $20$~dB, $40$~dB. Most importantly, we see that generally as $\epsilon$ increases, the average recovery error first drops and then increases. The error is small at the minimum. The best performance occurs for $\epsilon$ when the size of $(\phi, \lambda/2+\epsilon)$ admissible partition of $V'$ is relatively small. In such a case, we have a balance between the partition size and amount of nodes in each partition. In addition, the best range of $\epsilon$ is independent of the noise level.

The observation made above gives us a general guideline on how to choose $\epsilon$. One may first choose $\epsilon_1< \epsilon_2$ such that: at $\epsilon_1$, the number of partitions is $\approx K+K'$; while at $\epsilon_2$, the number of partitions is $\approx 1$. Then one may perform a binary search scheme within the interval $[\epsilon_1, \epsilon_2]$ to identify a range of $\epsilon$ so that the partition size is appropriate.

For illustration, we show in \figref{fig:gs_ex} examples of the recovery of continuous-time folded graph signals. \figref{fig:gs_ex}(a) shows an example with relatively good recovery while \figref{fig:gs_ex}(b) shows one with less perfect recovery. However, even in \figref{fig:gs_ex}(b), the error in relative difference between adjacent time slots in the recovered signals is still small as compared with the relative difference of the original signal.

\begin{figure}[!htbp]
	\footnotesize
	\centering
	\begin{minipage}[b]{.5\linewidth}
		\centering
		\centerline{\includegraphics[width=8.3cm, height=6cm]{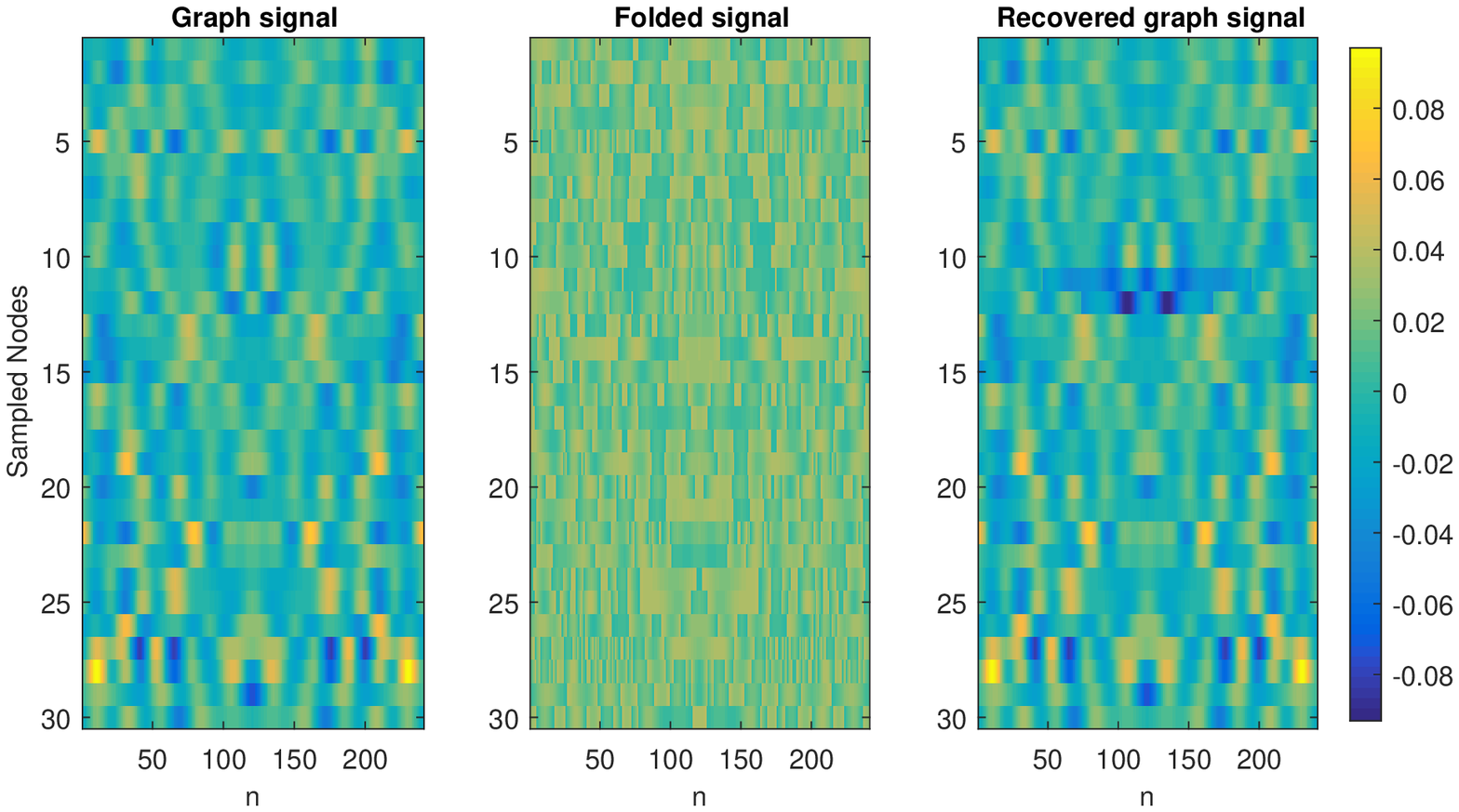}}
		\centerline{(a)}
	\end{minipage}%
	\begin{minipage}[b]{.5\linewidth}
		\centering
		\centerline{\includegraphics[width=8.3cm, height=6cm]{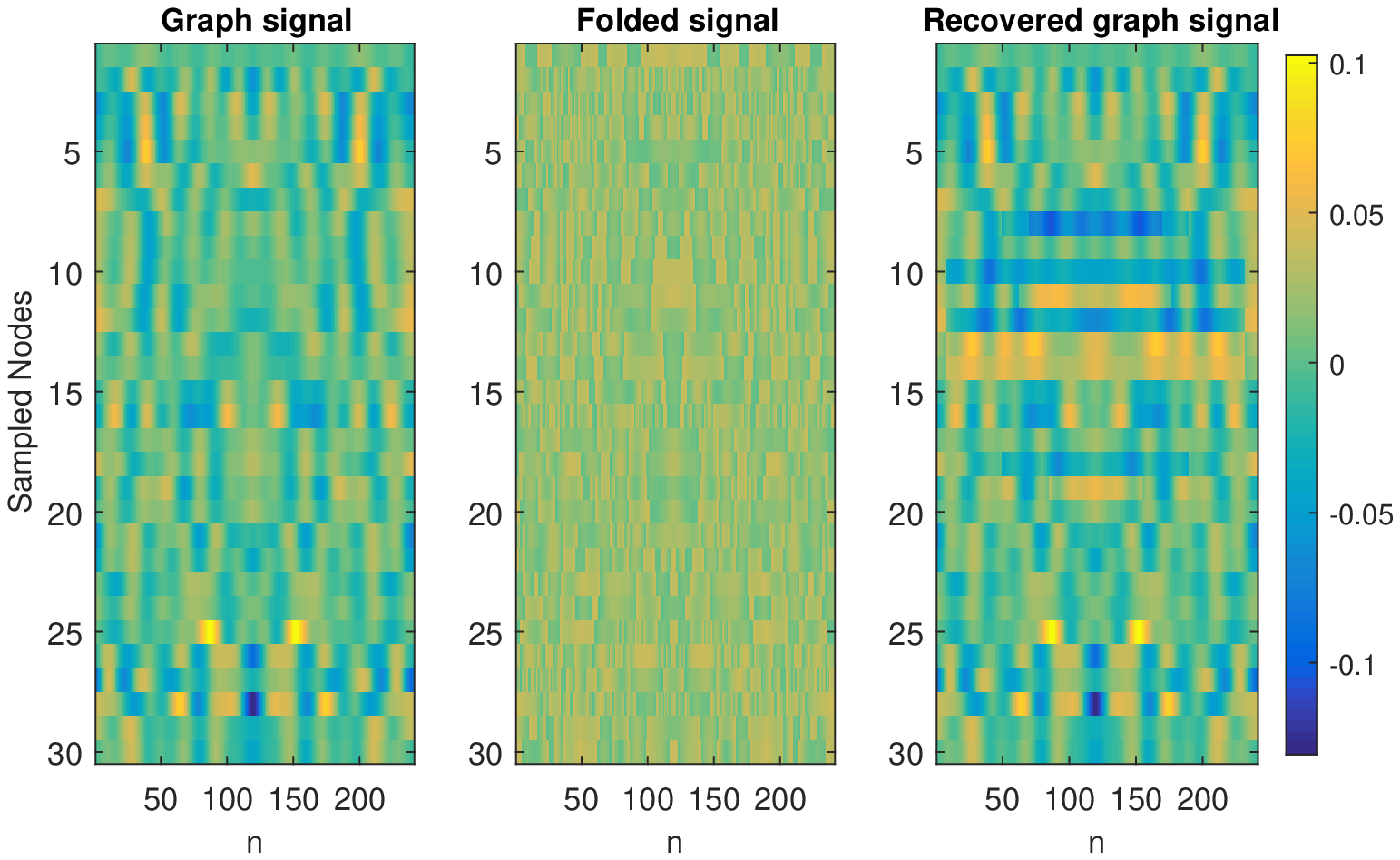}}
		\centerline{(b)}
	\end{minipage}
	\caption{The example heat maps of continuous-time folded graph signals: the original unfolded signals (left), and folded signals (middle) and the recovered signals (right). The horizontal axis is the time domain, while the vertical axis corresponds to the graph vertex domain.}
	\label{fig:gs_ex}
\end{figure}

\subsection{Folded images}

In this subsection, we present simulation results for folded image recovery. As there are no time component, we remove the integral factor in (\ref{eq:phi}) when performing folded image recovery. For each image, let $P$ be the total number of pixels. We express parameters such as $K,$ $K'$ as a factor of $P$ for convenience.  

\begin{figure}[!htbp]
	\footnotesize
	\centering
	\begin{minipage}[b]{.5\linewidth}
		\centering
		\centerline{\includegraphics[scale=0.6]{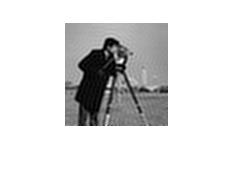}}
		\centerline{(a)}
	\end{minipage}%
	\begin{minipage}[b]{.5\linewidth}
		\centering
		\centerline{\includegraphics[scale=0.6]{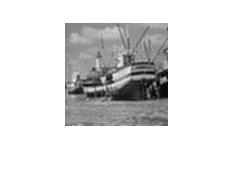}}
		\centerline{(b)}
	\end{minipage}
	\caption{Recovered images of the folded images in \figref{fig:5}.}\label{fig:7}
\end{figure}

For the toy examples in \figref{fig:5}, the pixel values are scaled within $[0,1]$ and the folding rate is taken to be $0.75$. \figref{fig:5}(a) has $K = 0.2P$, $K'=0.05P$. \figref{fig:5}(b) has $K=0.3P$, $K'=0.05P$. \figref{fig:5} looks darkened and noisy as we assign $0$ at unobserved pixels, though they are not used in the recovery procedure. Using our approach described in Section~\ref{sec:gsr}, the images are recovered perfectly as as shown in \figref{fig:7}.

Next, we consider the recovery of a mouse brain image using a self-reset CMOS sensor \cite{Tak16}. In this experiment, an image sensor was created by \cite{Tak16} and surgically implanted onto the brain surface of a mouse (see Fig.~4 of \cite{Tak16}). The image sensor is designed with a self-reset circuit to enable it to work with high dynamic range of light intensity and SNR. The reset count of each pixel was recorded so that the original pixel light intensity value can be recovered. We make use of the folded image created by \cite{Ayush2017}, which is based on the grayscale image in Fig.~7 of \cite{Tak16} to recover the original image. Our goal is to show that we can recover the original image without keeping track of the reset counts.

\begin{figure}
	\centering
	\includegraphics[scale=0.85]{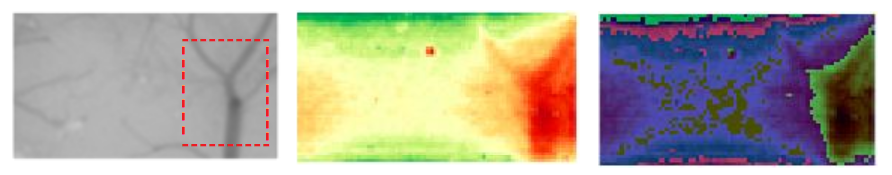}
	\caption{The mouse brain surface images. Left: the original grayscale image with highlighted key feature. Middle: the colored image. Right: the folded image of the colored version} \label{fig:6}
\end{figure}

The mouse brain surface shown on the left of \figref{fig:6} is obtained using a microscope with white light and is provided by \cite{Tak16}. The key feature captured by the image is the blood vessel highlighted within the dotted square. The colored version (middle image of \figref{fig:6}) is taken from \cite{Ayush2017} with bandlimit $K \approx 0.57P$. The blood vessel in the grayscale image is captured in the red region towards the right end. If the folding rate $\lambda=0.7$, we obtain the folded image (right of \figref{fig:6}). A large part of the image has non-zero folding numbers: $89\%,83\%, 8\%$ for the red, green and blue channels respectively. In the folded image, the blood vessel feature is lost due to the folding effect. 

\begin{figure}
	\centering
	\includegraphics[scale=0.82]{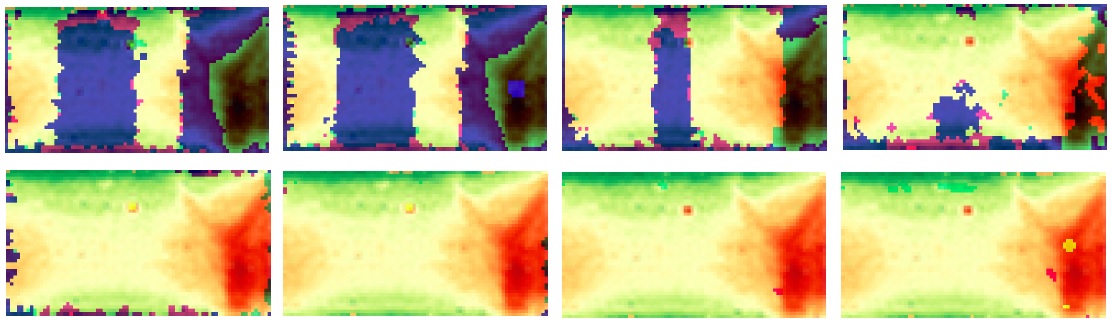}
	\caption{The recovery results of the folded mouse brain surface image. Parameters from left to right and top to bottom: $\epsilon = 14\lambda, 16\lambda, \ldots, 30\lambda$. Size of a $(\phi, \lambda/2+\epsilon)$-admissible partition: $0.33P, 0.34P, 0.21P, 0.20P, 0.14P, 0.076P, 0.054P, 0.032P$. Average recovery errors of the RGB channels: $36.7\%, 37.0\%, 18.0\%, 11.0\%, 2.61\%, 0.41\%, 0.13\%, 0.74\%$.} \label{fig:4}
\end{figure}

To perform the recovery, we make use of all the pixels, i.e., $K+K'=P$. In addition, we assume unfolded observations are made at $\approx 0.05P$ pixels. We test extensively with different $\epsilon$ values in forming $(\phi, \lambda/2+\epsilon)$-admissible partitions of $V$.  The recovery results are shown in \figref{fig:4}. We see that as $\epsilon$ increases the size of a $(\phi, \lambda/2+\epsilon)$-admissible partition decreases, while the recovery performance increases initially. Moreover, when the partition size becomes too small, the performance starts to drop again. For all the four recovered images at the bottom, the blood vessel feature can be seen clearly. 

\begin{figure}[!htbp]
	\footnotesize
	\centering
	\begin{minipage}[b]{.5\linewidth}
		\centering
		\centerline{\includegraphics[width=7.5cm, height=7cm]{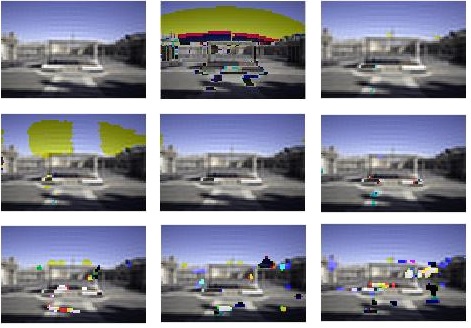}}
		\centerline{(a)}
	\end{minipage}%
	\begin{minipage}[b]{.5\linewidth}
		\centering
		\centerline{\includegraphics[width=7.5cm, height=7cm]{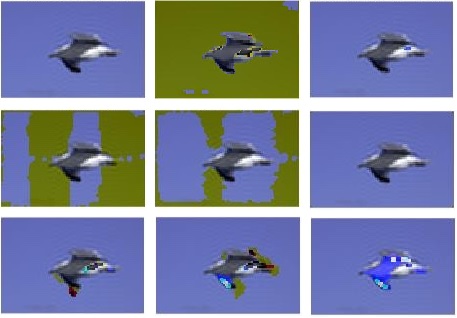}}
		\centerline{(b)}
	\end{minipage}
	\caption{Row 1: the original image, the folded image, and the recovered image by majority voting. Rows 2 and 3: recovered images with $\epsilon = 16\lambda, 18\lambda, 20\lambda, 22\lambda, 24\lambda, 26\lambda$, respectively.} \label{fig:9}
\end{figure}

We perform two additional sets of experiments on (compressed versions of) colored photo images.\footnote{Photos with ID 0030 and 0044 of http://data.csail.mit.edu/graphics/fivek/.} Let $p_{\max}$ be the maximum pixel value. For both images, we choose $\lambda = 0.7p_{\max}$. The folded images have severe color distortion (\figref{fig:9}). We show the recovery results with $\epsilon = 16\lambda, 18\lambda, 20\lambda, 22\lambda, 24\lambda, 26\lambda$ in \figref{fig:9}. 

As observed before in the other experiments, for each image, the recovery performance first improves as $\epsilon$ increases and then drops. We may fuse the results for the $6$ recovered images by taking a simple \emph{majority voting} of the folding numbers. The fused images are also shown in \figref{fig:9}, and we see that they give good recovery of the original images.

\section{Conclusion}\label{sec:c}

In this paper, we study a spatio-temporal sampling approach for graph signals while considering a practical scenario of modulo-based sampling using self-reset ADC for high dynamic range signals. A theoretical graph sampling rate for recovery from folded signals has been provided, which requires only the Nyquist-Shannon rate in the time direction. However, unconditional perfect recovery requires integer programming, which is intractable for large graphs. Under certain smoothness assumptions on the signals, we propose a new graph sampling scheme that minimize certain complexity measure introduced in the paper. Simulation results have been provided to demonstrate the performance of the proposed approach. 

For future work, it is of interest to investigate improvement of our approach to make it more robust when the signal has large components in the high frequency regime.

\appendices

\section{Graphs with random edge weights} \label{app:gre}

In this appendix, we consider graphs with random edge weights, and demonstrate that several conditions being assumed in Theorem~\ref{thm:1} and other places of the paper are common in this probabilistic setting.

Let $G=(V,E)$ be an undirected graph of size $n=|V|$. The edges in $E$ are used to denote topological connections between pairs of nodes, and does not carry any metric information. We assign random weights to edges in $E$. More precisely, for each $e \in E$, let its weight $a_e$ be a positive number chosen randomly according to a probability distribution absolutely continuous w.r.t.\ the Lebesgue measure. The weights are drawn independently for different edges. $G$ coupled with $A$, the associated weighted adjacency matrix, is called a \emph{weighted model} of $G$. As the explicit choice of the probability distributions does not play a role in the discussion, we suppress them in the discussion.    

We say that a property $P$ \emph{holds generally for $G$} if it holds for weighted models of $G$ with probability one. A obviously weaker version is that $P$ holds for at least one weighted model of $G$. We are interested in a property that holds generally for $G$ as long as it holds for at least one weighted model. 

For a weighted model $(G,A)$, let $L$ be its weighted Laplacian matrix. Fix a positive integer $K\leq n$, a subset $S \subset V$ of size $K$ and a node $u \in V\backslash S$ in case $K<n$. Arrange the eigenvalues $\lambda_1\leq \ldots \leq \lambda_n$ of $L$ in ascending order and let $w_i$ be an eigenvector corresponding to $\lambda_i$ such that $\{w_1,\ldots, w_n\}$ forms an orthonormal basis, whose existence is guaranteed by the spectral theorem. The collection of the first $K$ basis element is denoted by $W=\{w_1,\ldots, w_K\}$. The matrix $W_S$ and vector $W_u$ are defined as in Section~\ref{sec:fgs}. 

We consider the following list of related properties:
\begin{enumerate}[(a)]
	\item $P_0:$ $\lambda_1,\ldots, \lambda_n$ are all distinct.
	
	\item $P_1:$ $W_S$ is invertible.
	
	$Q_1:$ For any $K$ distinct orthonormal eigenvectors $W' = \{w_1',\ldots, w_K'\} \subset \{w_1,\ldots, w_n\}$ of $L$, $W'_S$ is invertible. 
	
	\item $P_2:$ the entries of $W_uW_S^{-1}$ are linearly independent over $\mathbb{Z}$.
	
	$Q_2:$ For any $K$ distinct orthonormal eigenvectors $W' = \{w_1',\ldots, w_K'\} \subset \{w_1,\ldots, w_n\}$ of $L$, the entries of $W'_u{W'_S}^{-1}$ are linearly independent over $\mathbb{Z}$. 
	
	\item $P_3:$ $W_uW_S^{-1}$ contains an irrational entry.
	
	$Q_3:$ For any $K$ distinct orthonormal eigenvectors $W' = \{w_1',\ldots, w_K'\} \subset \{w_1,\ldots, w_n\}$ of $L$, $W'_u{W'_S}^{-1}$ contains an irrational entry. 
	
	\item $P_4:$ the union of $1$ and entries of $W_uW_S^{-1}$ are linearly independent over $\mathbb{Z}$.
	
	$Q_4:$ For any $K$ distinct orthonormal eigenvectors $W' = \{w_1',\ldots, w_K'\} \subset \{w_1,\ldots, w_n\}$ of $L$, the union of $1$ and entries of $W'_u{W'_S}^{-1}$ are linearly independent over $\mathbb{Z}$.
\end{enumerate}

In $P_2, P_3, P_4$ (resp.\ $Q_2, Q_3, Q_4$) we implicitly assume $P_1$ (resp.\ $Q_1$) holds correspondingly. It is clear that if all of these properties holds generally for $G$ separately, a combination of them also holds generally for $G$. Moreover, if $Q_i$ holds generally for $G$, as a special case, so does $P_i, i=1,2,3,4$. On the other hand, it is possible to have $G$ such that some $P_i$ does not hold for any model of $G$, e.g., $P_0$ does not hold if $G$ is disconnected. 

To prepare for the discussions, we briefly discuss some key concepts and ideas. Details can be found in standard text books on differentiable manifolds, e.g., \cite{War83}. Recall that a differentiable manifold (abbreviated as manifold) $X$ of dimension $n$ is a topological space such that: at every point of $x\in X$, there is an open neighborhood $U_x$ identified with an open subset of $\mathbb{R}^n$ via a map $f_x$. Notion of differentiability can thus be defined on $f_x(U_x)$. If $U_x$ and $U_y$ intersect, then they are glued together via the transition function $\psi_{x,y}= f_y\circ f_x^{-1}: f_x(U_x\cap U_y) \to f_y(U_x\cap U_y)$. These $\psi_{x,y}$ are differentiable. The tangent space $T_x$ at any point $x\in X$ is an $n$-dimensional vector space. A differentiable map between two manifolds $X \to Y$ is called a diffeomorphism if it has a differentiable inverse.

An $m$-dimensional manifold $Y$ is a submanifold of $X$ if there is a one-to-one map $f: Y \to X$ that induces a non-singular linear transformations between tangent spaces at every $y \in Y$. We call $n-m$ the co-dimension of $Y$. 

A general strategy for the proofs of results in the Appendix goes as follows. We shall first construct a manifold $X$ parametrized by graph edge weights, usually such an $X$ carries the Lebesgue measure. If $Y$ is a closed submanifold with co-dimension at least $1$, detectable by looking at the derivative (or Jacobian) of the defining conditions, then it has measure $0$. Or if $Y$ is the locus of analytic functions (e.g., polynomials), then it is either all of $X$ or has measure $0$. For example, the locus of a single non-zero polynomial is a finite union of submanifolds each of co-dimension at least $1$, and the locus has $0$ Lebesgue measure.

We separate the discussion of $P_0$ from the rest, as $W$ may not be unique if $P_0$ fails. 

\begin{Lemma} \label{lem:pt}
	If $P_0$ holds for at least one weighted model, then $P_0$ holds generally for $G$.
\end{Lemma}
\begin{IEEEproof}
For a square matrix $M$, let $P_M(x) = \det(xI-M)$ be its characteristic polynomial, where $I$ is the identity matrix of the same size. 

For a weighted model $(G,A)$, the weighted Laplacian matrix $L$ is parametrized by the strict upper triangular entries of $A$, i.e., the random edge weights $\{a_e : e\in E\}$. Thus, $P_L(x)$ is a polynomial on the variables $\{a_e : e\in E\} \cup \{x\} \subset \mathbb{R}^{|E|+1}$. For a given $\{a_e : e\in E\}$, $P_0$ fails if and only if there exists an $x\in\Real$ such that $P_L(x) = 0$ and $P_L'(x) = \ud P_L(x)/\ud x = 0$ are satisfied simultaneously. Denote the simultaneous solution set $\{a_e : e\in E\} \cup \{x\}$ to be $X$. 

Let $\mathbb{R}^{|E|}_{\geq 0}$ be the coordinates of the variables $\{a_e : e \in E\}$. If $X \cap \mathbb{R}^{|E|}_{\geq 0}$ is of co-dimension at least $1$ in $\mathbb{R}^{|E|}_{\geq 0}$, then it has zero Lebesgue measure and hence $P_0$ holds generally. Otherwise, $X \cap \mathbb{R}^{|E|}_{\geq 0} = \mathbb{R}^{|E|}_{\geq 0}$. This is impossible as we assume that $P_0$ holds for at least one weighted model $(G,A)$. The lemma is now proved.
\end{IEEEproof}

Now, we work under the assumption that $P_0$ holds generally for $G$. As a consequence, we do not have any ambiguity in defining $W$, $W_S$, and $W_u$. 

\begin{Proposition} \label{prop:sph}
	Suppose $P_0$ holds generally for $G$. For any $i=1,2,3,4$, if $Q_i$ holds for at least one weighted model, then $Q_i$ holds generally for $G$. In particular, $P_i$ holds generally for $G$.
\end{Proposition}

\begin{IEEEproof}
Let $X\cong \mathbb{R}_{\geq 0}^{|E|}$ be the manifold parametrized by the strict upper triangular entries of $A$. Denote the Lebesgue measure on $X$ by $\mu$. As we assume that $P_0$ holds generally for $G$, there is an open submanifold $X_0$ consisting of weighted models whose Laplacian $L$ has no repeated eigenvalues and whose complement is of co-dimension at least $1$.

On the submanifold $X_0$, the Schur decomposition \cite{Die99} defines a smooth map 
\begin{align*}
\psi: X_0 \to \mathbb{R}^{n^2} \times \mathbb{R}^{n}, A \mapsto (P_A,D_A),
\end{align*}
where $L=P_A D_A P_A^{-1}$ is the unique Schur (orthonormal) decomposition of $L$. One should take note that the diagonal entries of $D_A$ does not necessarily correspond to the ascending arrangement of the eigenvalues of $A$. Let $Y$ be the image of $\psi$. Then, $\psi: X_0 \to Y$ is a diffeomorphism, whose inverse is given by $(P,D) \mapsto PDP^{-1}$, which is a polynomial. 

\emph{$Q_1$:} For $K$ distinct indices $I$ of $\{1,\ldots,n\}$, let $W_I = \{w_i : i\in I\}$ ($W_I$ also denotes the matrix whose columns are $w_i,i\in I$) and $W_{I,S}$ be the submatrix of $W_I$ consisting of rows indexed by $S$. Its determinant $p_I$ is a  polynomial on the coordinates of $Y\subset \mathbb{R}^{n^2}$. If $p_I$ is non-vanishing on $Y$, then we are done. Otherwise, as $Q_1$ holds for at least one weighted model, $p_I$ is not the constant $0$ polynomial. Therefore, the locus $Z_I$ of points of $p_I=0$ is a finite union of submanifolds of co-dimension at least $1$ in $Y$, and consequently $\mu(\psi^{-1}(Z_I))=0$. Define $Y_1 = Y\backslash(\cup_I Z_I)$. It follows immediately that the inverse image of the complement of $Y_1$ has zero Lebesgue measure. On the other hand, for any weighted model does not satisfy $Q_1$, its Laplacian matrix belongs to $X_0\backslash \psi^{-1}(Y_1)$. Therefore, $Q_1$ holds generally for $G$.  

\emph{$Q_2$:} Let $a=(a_1)_{1\leq i\leq K} \in \mathbb{Z}^K\backslash \{0\}$ be a nonzero (column) vector and $I$ be $K$ distinct indices $I$ of $\{1,\ldots,n\}$. Denote the subset of $Y_1$ such that the polynomial $\det(W_{I,S})W_{I,u}W_{I,S}^{-1}a = 0$ by $Z_{I, a}$. Suppose $Q_2$ holds for some models of $G$. As above, $Z_{I,a}$ is a submanifold of co-dimension at least $1$. Hence, $\psi^{-1}(Z_{I,a})$ has zero Lebesgue measure. As the collection $\psi(\{Z_{I,a}\})$ is countable, their union has zero measure. Therefore, $Q_2$ holds generally for $G$.

The remaining cases are similar, and we shall be brief.

\emph{$Q_3$:} Let $r$ be a rational number, $I$ be $K$ distinct indices of $\{1,\ldots, n\}$, and $\{e_1,\ldots, e_K\}$ be the standard basis of $\mathbb{R}^K$. Denote the subset of $Y_1$ such that $\det(W_{I,S})(W_{I,u}W_{I,S}^{-1} e_i -r)=0$ by $Z_{I,r,i}$. If $Q_3$ holds for some models of $G$, $\mu(\psi^{-1}(Z_{I,r,i}))=0$ as above. On the other hand, $Q_3$ holds for a model of $G$ if it does not belong to any of $\psi^{-1}(Z_{I,r,i})$, which is a countable collection. Therefore, $Q_3$ holds generally for $G$.

\emph{$Q_4$:} Let $a=(a_1)_{1\leq i\leq K} \in \mathbb{Z}^K\backslash \{0\}$ be a nonzero (column) vector, $I$ be $K$ distinct indices $I$ of $\{1,\ldots,n\}$ and $b \in \mathbb{Z}$. Denote the subset of $Y_1$ such that $\det(W_{I,S})(W_{I,u}W_{I,S}^{-1}a + b)=0$ by $Z_{I,a,b}$. If $Q_4$ holds for some models of $G$, $\mu(\psi^{-1}(Z_{I,a,b}))=0$ as above. On the other hand, $Q_4$ holds for a model of $G$ if it does not belong to any of $\psi^{-1}(Z_{I,a,b})$, which is a countable collection. Therefore, $Q_4$ holds generally for $G$.
\end{IEEEproof}

We remark here that the same argument also works without any changes if we are able to find a model with possibly negative edge weights. This relaxation of condition can be convenient in some arguments.

\begin{Example}
We shall use general, abstract but non-constructive arguments below, although in certain special cases, one can also perform explicit calculations. 	
	
\begin{enumerate}[(a)]	
\item \label{it:lgb} Let $G$ be the path graph on $n$ nodes for $n\geq 2$. We claim that $P_0$ holds generally for $G$. Let $L_n$ be the Laplacian matrix of $G$ with weight $1$ for each edge and $K_n$ be the $n\times n$ matrix obtained from $L_n$ by replacing the $(n,n)$ entry of $L_n$ (which is $1$ in $L_n$) by $2$.
 
We first show by induction that: (1) $K_n$ does not have repeated eigenvalues; and (2) $K_n$ and $K_{n+1}$ do not have common eigenvalues. It is straightforward to verify directly that both claims hold for $K_2$ and $K_3$. Let $\lambda$ be an eigenvalue of $K_{n-1}$, i.e., $\det(K_{n-1}-\lambda I_{n-1})=0$, and $I_n$ be the $n\times n$  identity matrix, $n\geq 4$. By the induction hypothesis, $\det(K_{n-2}-\lambda I_{n-2}) \neq 0$. Apply the Laplace formula of matrix determinant (expanded w.r.t.\ the last row), one obtains: 
\begin{align*}
|\det(K_n-\lambda I_n)| & = |(2-\lambda)\det(K_{n-1}-\lambda I_{n-1}) - (-1)\cdot \det(K_{n-2}-\lambda I_{n-2})| \\
& = |\det(K_{n-2}-\lambda I_{n-2})| \neq 0.  
\end{align*}
 
Therefore, $\lambda$ is not an eigenvalue of $K_n$ and we have proved claim (2). Let $\lambda_1 \leq \ldots \leq \lambda_n$ be the eigenvalues of $K_n$ and $\delta_1 < \ldots < \delta_{n-1}$ be the distinct eigenvalues (by the induction hypothesis) of $K_{n-1}$. As the top-left block of $K_n$ is $K_{n-1}$, we may invoke the Cauchy interlacing theorem (\cite{Hwa04}) to obtain
\begin{align} \label{eq:lld}
\lambda_1\leq \delta_1 \leq \lambda_2 \leq  \ldots \leq \lambda_{n-1} \leq \delta_{n-1} \leq \lambda_n.
\end{align}
However, as we have proved that $K_n$ and $K_{n-1}$ have distinct eigenvalues, the inequalities in (\ref{eq:lld}) are all strict. Therefore, $\lambda_1,\ldots,\lambda_n$ are distinct and this proves claim (1). Using the same argument, one can show that each $L_n$ has distinct eigenvalues based on the fact that each $K_n$ has distinct eigenvalues. Therefore, $P_0$ holds generally for $G$ by Lemma~\ref{lem:pt}.

As a consequence, $P_0$ holds generally for any $G$ that contains a path connecting all the nodes.
 
\item Let $G$ be the complete (simple) graph on $n$ nodes for $n> 2$. We claim that each $P_i, i=0,\ldots, 4$ holds generally for $G$. First of all, as the complete graph contains a path of all the nodes, $P_0$ holds generally for $G$ by \ref{it:lgb}.

For the other statements, we want to apply Proposition~\ref{prop:sph} to produce example for each individual case. As mentioned above, we allow models with negative edge weights. The space of all such models $X$ can thus be identified with $\mathbb{R}^{n(n-1)/2}$. The Schur decomposition of the Laplacian matrix defines a map $\psi: X \to O(n)$, where $O(n)$ is the group of orthogonal matrices. On the other hand, for $P \in O(n)$, it is in the image of $\psi$ if and only if there is a non-zero diagonal matrix $D$ such that $PDP^{T}{\bf 1} = 0$, where ${\bf 1}$ is the all $1$ column vector. This holds if and only if the sum of at least one column of $P$ is $0$. Therefore, $\psi(X)$ is the union of loci of $z\in O(n)$ satisfying the linear relation $l_i(z) = 0$, where $l_i(z)$ is the sum of the $i$-th column.

Let $x$ be a model in $X$ such that its Laplacian matrix $L_x$ does not have repeated eigenvalues. As in the proof of Lemma~\ref{lem:pt}, there is an open neighborhood $U_x$ of $x$ such that
\begin{enumerate}[(i)]
	\item For each $y\in U_x$, $L_y$ does not have repeated eigenvalues.
	\item $\psi(U_x)$ is an open subset of $\psi(X)$.
\end{enumerate}      

For demonstration purpose, we explain $Q_2$ holds for some models. The same reasoning applies to $Q_1, Q_3, Q_4$ as well. It can be verified (by computing on test vectors) that for any non-zero vector of integers $a \in \mathbb{Z}^K$, the polynomial (on the matrix entries) $\det(M_S)M_uM_S^{-1}a$ does not contain any of $l_i$ as a factor, i.e., the loci of $l_i$ and $\det(M_S)M_uM_S^{-1}a$ intersects transversally in $O(n)$. Therefore, $\psi(U_x)$, which is an open subset of $O(n)$ satisfying at least one $l_i$, contain enough points to avoid the countable union of $\det(M_S)M_uM_S^{-1}a=0$. Therefore, $Q_2$ holds for some models, and consequently, $P_2$ holds generally for $G$.
\end{enumerate}
\end{Example}

\bibliographystyle{IEEEtran}
\bibliography{IEEEabrv,StringDefinitions,reference}
\end{document}